\documentclass[twocolumn]{aastex6}

\newcommand{\avg}[1]{\langle #1 \rangle}
\usepackage{amsmath}
\usepackage{booktabs}
\usepackage{lipsum}
\usepackage{enumitem}

\shorttitle{Acceleration of solar wind particles at CME shocks}
\shortauthors{Prinsloo et al.}

\begin{document}

\title{Acceleration of solar wind particles by traveling interplanetary shocks}


\author{P. L. Prinsloo\altaffilmark{1} and R. D. Strauss\altaffilmark{2}}
\affil{Centre for Space Research, North-West University, 2520 Potchefstroom, South Africa}
\and
\author{J. A. le Roux}
\affil{Department of Space Science, University of Alabama in Huntsville, Huntsville, AL 35805, USA\\
Centre for Space Plasma and Aeronomic Research, University of Alabama in Huntsville, Huntsville, AL 35899, USA}


\altaffiltext{1}{email: drageoprinsloo@gmail.com}
\altaffiltext{2}{email: dutoit.strauss@nwu.ac.za}

\begin{abstract}

The acceleration of thermal solar wind protons at spherical interplanetary shocks driven by coronal mass ejections is investigated.
The solar wind velocity distribution is represented using $\kappa$-functions, which are transformed in response to simulated shock transitions in the fixed-frame flow speed, plasma number density, and temperature.
These heated solar wind distributions are specified as source spectra at the shock from which particles with sufficient energy can be injected into the diffusive shock acceleration process.
It is shown that for shock-accelerated spectra to display the classically expected power-law indices associated with the compression ratio, diffusion length scales must exceed the width of the compression region.
The maximum attainable energies of shock-accelerated spectra are found to be limited by the transit times of interplanetary shocks, while spectra may be accelerated to higher energies in the presence of higher levels of magnetic turbulence or at faster-moving shocks.
Indeed, simulations suggest fast-moving shocks are more likely to produce very high-energy particles, while strong shocks, associated with harder shock-accelerated spectra, are linked to higher intensities of energetic particles.
The prior heating of the solar wind distribution is found to complement shock acceleration in reproducing the intensities of typical energetic storm particle events, especially where injection energies are high.
Moreover, simulations of $\sim$0.2 to 1 MeV proton intensities are presented that naturally reproduce the observed flat energy spectra prior to shock passages.
Energetic particles accelerated from the solar wind, aided by its prior heating, are shown to contribute substantially to intensities during energetic storm particle events.  

\end{abstract}

\keywords{acceleration of particles  
--- solar wind 
--- Sun: heliosphere
--- shocks}

\section{Introduction} \label{sec:intro}

Energetic particle enhancements observed at Earth are often associated with the passage of interplanetary (IP) shocks, driven by e.g. coronal mass ejections (CMEs), during energetic storm particle (ESP) events \citep{Bryantetal1962,LarioDecker2002,Hoetal2009,HuttunenHeikinmaaValtonen2009,Makelaetal2011}.
These enhancements are thought to occur as a result of particle acceleration at the shock, for which diffusive shock acceleration \citep[DSA, ][]{Axfordetal1977,Krymsky1977,Bell1978a,BlandfordOstriker1978} continues to be regarded a viable mechanism, often coupled with self-generated turbulence in the form of magnetohydromagnetic waves \citep{Lee1983}. 
See also the relevant discussions of \cite{DesaiGiacalone2016} on ESPs and DSA.
More recently, attention has been drawn to the notable flattening of proton energy spectra in observations directly preceding the passage of IP shocks at Earth \citep{Larioetal2018}. 
This is attributed to follow as a result of the propagation of accelerated particles from the shock to the observer, but is not fully explained.
Moreover, while CME shock properties such as speed and compression ratio have been linked to the efficiency of particle acceleration \citep{Larioetal2005b,Giacalone2012}, shock obliquity \citep{Ellisonetal1995} and the nature of the seed particles \citep{Desaietal2006} have also been identified as important factors. 

It has been suggested that IP shocks can accelerate particles from thermal energies \citep[e.g.][]{Giacalone2005}. 
Indeed, for quasi-parallel shocks, where particles can repeatedly cross the shock front along magnetic field lines, injection energies are assumed to be small, and Maxwellian-like distributions are proposed to provide adequate seed particles for DSA \citep{Giacaloneetal1992,NeergaardParkerZank2012}. 
By contrast, injection energies are generally assumed to be high for perpendicular shocks, because the perpendicular diffusion coefficient is small for particles interacting resonantly with micro-scale turbulence.
However, the injection energy is considerably lowered when particles experience perpendicular diffusion along intermediate-scale meandering magnetic field lines.
The injection speed at perpendicular shocks can become comparable to those at parallel shocks, and DSA thereby more effective, if the amplitude of field line meandering is large enough \citep{Giacalone2005}.
On the other hand, the shock front itself may be rippled:
while an idealized shock propagating through an approximately radial magnetic field may vary from being quasi-perpendicular at the CME flanks to quasi-parallel near the nose, realistically, local geometries may resemble any obliquity \citep[see][]{KleinDalla2017}.

Aside from conducive shock geometry, injection into the DSA process is facilitated by the formation of non-thermal seed particles \citep{NeergaardParkeretal2014,Zank2017}, which has also recently been investigated using kinetic hybrid simulations \citep{Capriolietal2015, Sundbergetal2016}.
Suprathermal tails are often observed in solar wind (SW) velocity distributions \citep{Collieretal1996,Maksimovicetal1997,Chotooetal2000,Qureshietal2003} and are considered conducive features for the injection of particles into DSA \citep{Desaietal2006,Kangetal2014}.
To parameterize these distributions, \cite{Vasyliunas1968} introduced the Kappa($\kappa$)- distribution function, which characterizes both the Maxwellian core and the suprathermal tail at higher energies.
Refer to \cite{PierrardLazar2010} and \cite{LivadiotisMcComas2013} for complete reviews on the theory and applications of $\kappa$-functions.    
These functions characterise SW distributions using only three parameters, namely, the $\kappa$-parameter, which relates to the spectral index of the high-energy tail, and the equivalent temperature and number density of the plasma or particle species considered \citep[][]{Formisanoetal1973,ChateauMeyerVernet1991}.
Of course, quantities such as the plasma density and temperature are observed to change during the passage of an IP shock \citep[e.g.][]{LarioDecker2002}, which also transforms the $\kappa$-function and thereby the properties of the potential DSA seed population it represents. 

In this study, two broad scientific questions are addressed, namely, how the prior heating of the SW distribution affects acceleration at IP shocks, and what role these shock-accelerated particles play in producing the spectral features observed during ESP events. 
The acceleration of SW particles is modelled by solving a set of stochastic differential equations (SDEs; introduced in Section \ref{sec:ModelSDEs}), equivalent to the \cite{Parker1965} transport equation, in spherical symmetry.
The applications are limited to halo CME shocks expanding radially at constant speeds.
In this spherically symmetrical scenario, the shock normal is radially aligned so that the shock obliquity can be approximated using the \cite{Parker1958} spiral angle.

The manner in which the shock passage affects the energy distribution of SW particles is investigated both before and after injection into the DSA process:
It is firstly considered how the initial SW distribution transforms in response to changes in the plasma properties (Section \ref{sec:SWheating}), and later how the resulting distribution affects the subsequent DSA process as a seed population (Section \ref{sec:SWtoESPs}).
The classical spectral characteristics of DSA for travelling shocks, and the comparative efficiency of DSA at fast and strong shocks, are revisited in Sections \ref{sec:spectral_features} and \ref{sec:strongvsfast}, respectively.
In addition to reproducing these more typical DSA features, the model is also applied to investigate the spectral flattening reported by \cite{Larioetal2018} ahead of IP shock passages (Section \ref{subsec:spec_evolution}), as well as to identify the original spatial distributions and energies of seed particles (Section \ref{subsec:accsites}) that would eventually contribute to intensities during ESP events.

\section{Modelling CME shock-induced changes in the solar wind} \label{sec:SWheating}
 
 To describe both the thermal and suprathermal velocity distributions observed in the SW, the $\kappa$-function is implemented in terms of particle speed $v$ as 
  \begin{equation} \label{eqn:kappafunc}
 f_{\kappa} = A_{\kappa}\left( 1+\frac{v^2}{\kappa v_{\kappa}^2} \right)^{-\kappa -1} \text{,} 
 \end{equation}
where $v_{\kappa}^2$ is the generalised thermal speed given by 
\begin{equation} \label{eqn:vthermal}
v_{\kappa}^2 = \left(2\kappa - 3 \right)\frac{k_b T}{\kappa m_p} \text{,}
\end{equation}
with $k_{B}$ the Boltzmann constant and $m_{p}$ the proton mass, and where
\begin{equation} \label{eqn:kappanormconst}
A_{\kappa}=\frac{n}{\left( \pi \kappa v_{\kappa}^2\right)^{3/2}} \frac{\Gamma\left( \kappa+1 \right)}{\Gamma\left(\kappa-\frac{1}{2} \right)} 
\end{equation}
  is the normalization constant obtained when setting $\int f_{\kappa} d^3 v = n$, with $n$ taken as the SW number density and where $\Gamma$ is the Gamma function. 
  This normalization is discussed in some detail in Appendix \ref{subsec:on_kappa_funcs}.  
    
  The $\kappa$-parameter is related to the power-law index of the suprathermal tail.
  Note that Eq. \ref{eqn:kappafunc} is defined for $\kappa > 3/2$, and that if $\kappa \rightarrow \infty$, it reduces to a Maxwellian.   
 For the correct interpretation of Eq. \ref{eqn:vthermal} and the associated temperature $T$,
it is instructive to consider the discussion by \cite{Hellbergetal2009}: 
 $v_{\kappa}$ is introduced by \cite{Vasyliunas1968} as the most probable particle speed, associated with the non-relativistic kinetic energy of $E_{\kappa} = m_{p}v_{\kappa}^2/2$. 
 Evaluating the second moment of $f_{\kappa}$ yields a total energy of $NE_{N} = \frac{3}{2}NE_{\kappa} \kappa/(\kappa-3/2)$, where $E_{N}$ and $N$ are the mean energy per particle and the total number of particles, respectively.
 Eq. \ref{eqn:vthermal} then follows upon the introduction of the plasma temperature $T$ \citep[originally by][]{Formisanoetal1973} through the invocation of the equipartition theorem, $E_{N}=\frac{3}{2}k_{B}T$, for a monatomic gas.
Although this temperature definition and the foregoing assumption of the equipartition of energy are not strictly valid for non-Maxwellian distributions, their use in this manner has become standard practice and is generally considered appropriate \citep[see][and the references therein]{Hellbergetal2009}.

\subsection{Changes in plasma properties across the shock} \label{subsec:plasmaprop}

\begin{figure}[tp]
\figurenum{1}
\centering
\includegraphics[scale=0.4]{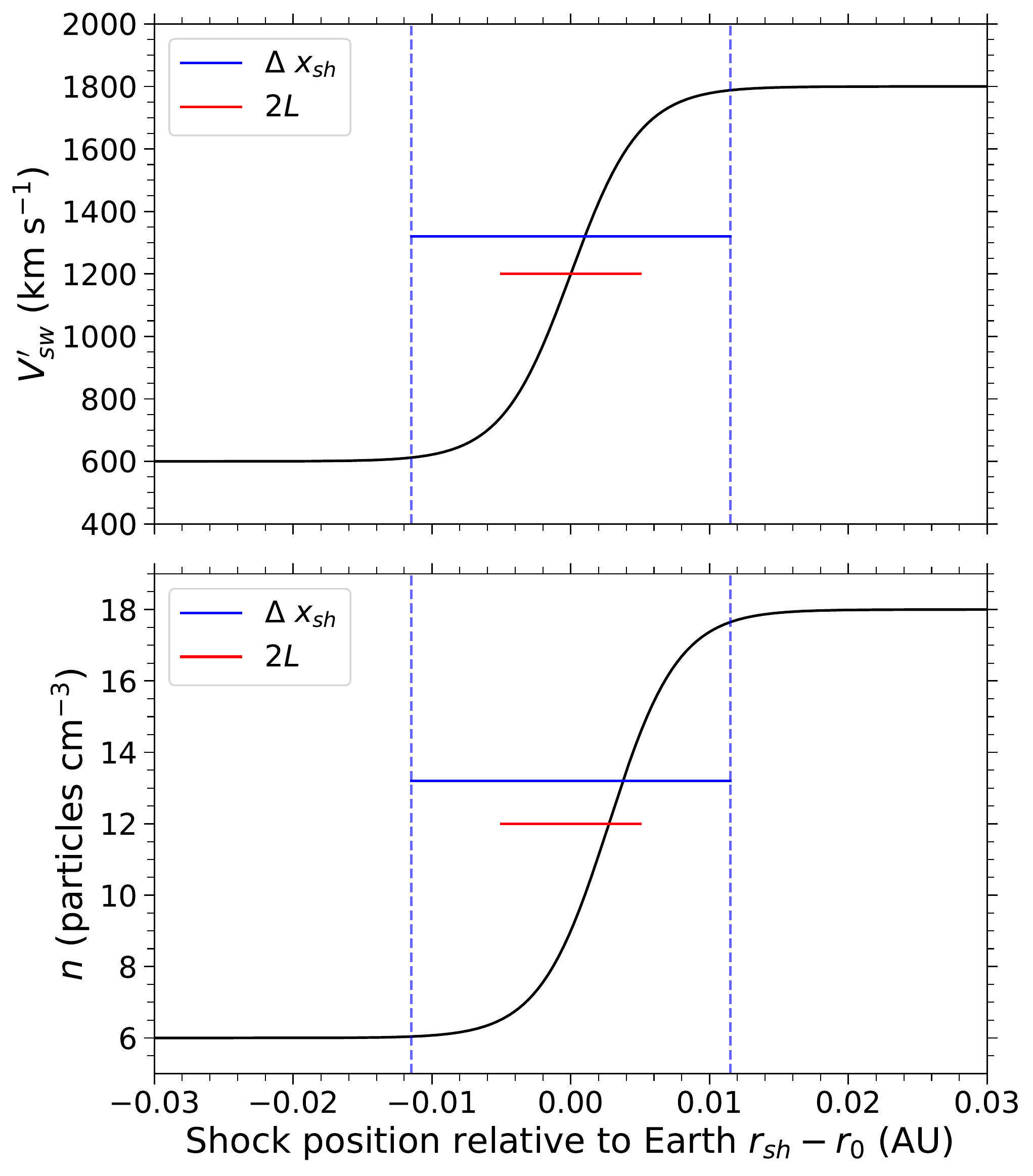}
\caption{Top: The transition of the fixed-frame SW flow speed $V_{sw}^{\prime}$, as viewed by an observer at the Earth, as a function of the shock position relative to that of the Earth ($r_{sh}-r_0$). $L\ (=0.005$ AU) and $\Delta x_{sh}$ are as defined for Eq. \ref{eqn:Vtransition} and Eq. \ref{eqn:L_relation}, respectively. Bottom: The SW number density profile corresponding to the flow-speed profile shown above.  \label{fig:flowdensity}}
\end{figure}

An objective of this study is to consider how the $\kappa$-function changes during the passage of an IP shock, and to implement this transformed distribution as a seed-particle spectrum for DSA.
To this end, while the $\kappa$-index is assumed constant across the shock, it is necessary to model the change in $n$ and $T$, upon which $f_{\kappa}$ depends \citep[see also][]{Livadiotis2015}.
For consistency, transitions across the shock are modelled to correspond to that of the flow speed as would be observed by e.g. a spacecraft in Earth's orbit.
The SW is consequently modelled to transition across the shock between up- and downstream flow speeds in the spacecraft (or fixed) frame, that is, $V_1^{\prime}$ and $V_2^{\prime}=\left(V_{sh}\left( s-1 \right)+V_1^{\prime}\right)/s$, respectively, according to
\begin{align} \label{eqn:Vtransition}
 V_{sw}^{\prime} &= \frac{1}{2s}\left(V_1^{\prime}\left(s+1\right)+V_{sh}\left(s-1\right)\right) \nonumber\\ 
 &-\frac{1}{2s}\left(\left(s-1\right)\left(V_{sh}-V_1^{\prime}\right) \tanh\left(\frac{r-r_{sh}}{L}\right)\right) \text{,} 
\end{align}
where $s$ is the shock compression ratio (or ratio of up- and downstream flow velocities in the shock frame), $V_{sh}$ is the shock speed, $r-r_{sh}$ is the radial position relative to the position of the shock $r_{sh}$, and $L$ is a characteristic length used to specify the broadness of this transition.
Figure \ref{fig:flowdensity} illustrates this transition for the reference parameters listed in Table \ref{tab:kappaparams}.
 Should $V_{sh}=0$, Eq. \ref{eqn:Vtransition} reduces to an expression \citep[used by e.g.][]{leRouxetal1996} for a stationary shock. 

Note furthermore that $L$ is not a direct measure of the shock width. Its extent can however be approximated from Eq. \ref{eqn:Vtransition} as $\triangle x_{sh} \approx 2r_{c}$, with
\begin{equation} \label{eqn:L_relation}
r_{c} = \tanh^{-1}\left( \frac{2sV_{c}-V_1^{\prime}(s+1)-V_{sh}(s-1)}{(1-s)(V_{sh}-V_1^{\prime})}\right) L \text{,}
\end{equation}
where a fraction $\mathcal{C} \in [0,0.5]$ is chosen such that $V_{c} = V_1^{\prime} + \mathcal{C}(V_2^{\prime}-V_1^{\prime})/2$  is the flow speed at some distance $r_c$ from the shock. Typically, to approximate the shock width, $\mathcal{C} =0.01$ so that $V_{c}\approx V_1^{\prime}$.  
It is illustrated in Figure \ref{fig:flowdensity} how the lengths $\triangle x_{sh}$, $L$, and the actual shock width compare.
 
 With the flow speed transition given, the change in number density $n$ across the shock follows from the continuity equation.
 Of course, the total factor by which the number density jumps across the shock must be equal to the compression ratio. 
 In terms of $V_{sw}^{\prime}$ it follows, in spherical coordinates, that   
\begin{equation} \label{eqn:ntransition}
n = n_0\frac{V_{sh}-V_1^{\prime}}{V_{sh}-V_{sw}^{\prime}} \left(\frac{r_0}{r}\right)^2 \text{,} 
\end{equation}  
where $n_0$ is the number density at some reference position such as the Earth ($r_0 = $ 1 AU), and where the flow speeds are transformed back to the shock frame, where their ratio is equal to $s$.
This transition is also shown in Figure \ref{fig:flowdensity}.

\begin{figure*}[tp]
\figurenum{2}
\centering
\includegraphics[scale=0.4]{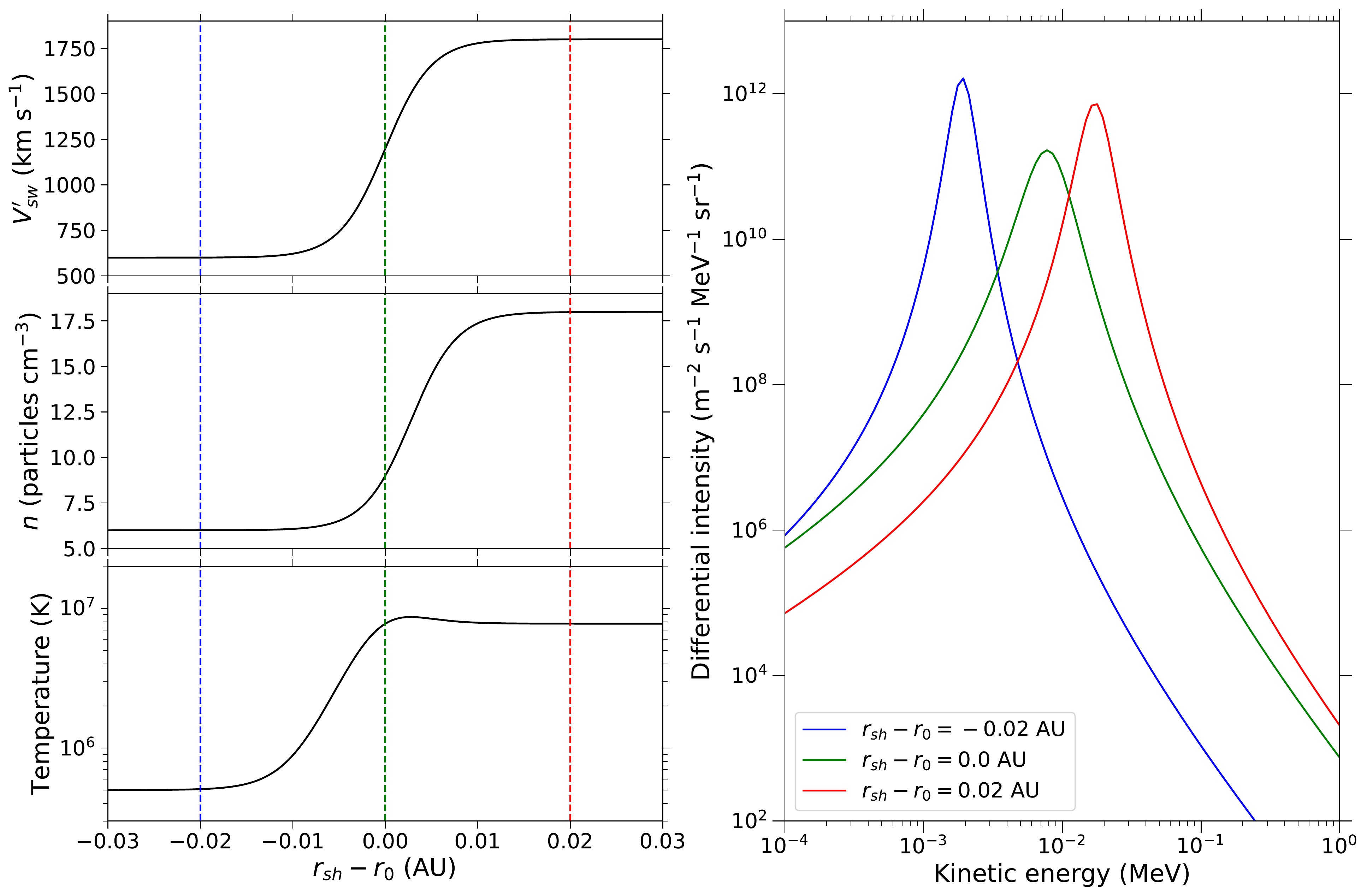}
\caption{Left: The transitions of the fixed-frame SW flow speed $V_{sw}^{\prime}$, number density $n$ and temperature $T$, as viewed by an observer at the Earth, as a function of the shock position relative to that of the Earth ($r_{sh}-r_0$).  Right: The heating of the SW energy distribution at the Earth in response to the shock transitions of plasma parameters as shown on the left. The curves represent SW distributions at different stages of the shock passage corresponding with the values of $r_{sh}-r_0$ indicated using vertical lines of similar colors on the left. \label{fig:plasmakappas}}
\end{figure*}

To estimate the jump in temperature across the shock, instead of implementing the relevant hydromagnetic Rankine-Hugoniot jump condition, a simpler approach capturing the essential physics is used.
The magnetic energy is expected to be much smaller than the thermal energy in the flow-dominated region considered in this study.
It is therefore assumed that the change in kinetic energy density $d\mathcal{E}_{sw}$ of the bulk solar wind flow, as a function of position through the shock, is converted to thermal energy. That is,
\begin{equation} \label{eqn:dE_transition}
d\mathcal{E}_{sw} = \frac{1}{2} m_p \left( n_0 \left( V_1^{\prime} - V_{sh}\right)^2 -  n \left( V_{sw}^{\prime} - V_{sh}\right)^2 \right) \text{,}
\end{equation} 
which is related through the equipartition theorem to the change in temperature $dT_{sw}$ by
\begin{equation} \label{eqn:dT_transition}
dT_{sw} = \frac{2}{\mathcal{N}_{df}} \frac{d\mathcal{E}_{sw}}{n k_b} \text{,} 
\end{equation}
where $\mathcal{N}_{df}=2/(\gamma_c-1)$ is the number of degrees of freedom and $\gamma_c$ is the ratio of specific heats. 
$dT_{sw}$ is subsequently added to the unshocked temperature value, that is, $T = T_0 +dT$, where
 \begin{equation}\label{eqn:T_unshocked}
 T_0 = T_e \left( r/r_0 \right)^{2\left(1-\gamma_c\right)}
 \end{equation} 
and $T_e$ is the plasma temperature at the Earth. 
For a monatomic gas for which $\gamma_c=5/3$ it follows that $T_0\propto r^{-4/3}$. 
It is assumed that the temperature gain during the shock passage is not significant enough to disassociate molecules, and hence $\gamma_c$ remains constant as for a calorically ideal gas.
Furthermore, given the multi-species composition of the SW plasma, $\gamma_{c}$ may deviate from the monatomic value of $5/3$.  
It is noted that while the hydromagnetic treatment of the shock jump conditions may break down in the transition region itself, it approximates the up- and downstream quantities adequately.
Since DSA requires particles to interact with the shock on length scales exceeding its width \citep[e.g.][]{JonesEllison1991,KruellsAchterberg1994}, this description is sufficient for the current study. See also Section \ref{subsec:fractional_compression}.

\begin{table}
\centering
\begin{tabular}{ccrl}
\toprule
$V_{sw}^{\prime}$: & $V_1^{\prime}$ & 600 & km s$^{-1}$ \\
 & $V_{sh}$ & 2400 & km s$^{-1}$ \\
 & $s$ & 3.0 & \\
 & $L$ & 0.005 & AU \\
$n$: & $n_0$ & 6.0 & cm$^{-3}$ \\
$T$: & $T_e$ & $5 \times 10^{5}$ & K \\
 & $\gamma_c$ & 3.5/3 & \\
$f_{\kappa}$: & $\kappa$ & 2.25 & \\
\bottomrule
\end{tabular}
\caption{Reference configuration of the parameters needed to define the fixed-frame flow speed $V_{sw}^{\prime}$, number density $n$, temperature $T$ and $\kappa$-function as discussed in Section \ref{subsec:plasmaprop}.}
\label{tab:kappaparams}
\end{table}

The shock transitions of $V_{sw}^{\prime}$, $n$ and $T$ are displayed in Figure \ref{fig:plasmakappas} as viewed by an observer at the position of the Earth. The parameters used to define these quantities are listed in Table \ref{tab:kappaparams} and are mostly informed by observations of CMEs during the 2003 Halloween epoch \citep{Skougetal2004,Richardsonetal2005,Gopalswamyetal2005,Larioetal2005a,Wuetal2005}. 
In particular, SW flow speeds in excess of 1800 km s$^{-1}$ \citep{Skougetal2004} and shock speeds of $\sim$2400 km s$^{-1}$ \citep{Gopalswamyetal2005,Wuetal2005} are noted.
Since density measurements during some larger events were uncertain \citep[e.g.][]{Skougetal2004,Wuetal2005}, the compression ratio for such events can be calculated using the ratios of shock-frame flow velocities.
For example, $s = (V_{sh}-V_1^{\prime})/(V_{sh}-V_{2}^{\prime})=3$ for $V_{sh} =$ 2400 km s$^{-1}$, $V_1^{\prime}=$ 600 km s$^{-1}$ and $V_{2}^{\prime}=$ 1800 km s$^{-1}$.   

\subsection{Evolution of the $\kappa$-function during the shock passage} \label{subsec:kappa_evolution}

Aside from $\kappa$, which is kept fixed, $f_{\kappa}$ is shown above to depend on parameters $n$ and $T$, both of which change during the shock passage.
Additionally, to illustrate how the SW distribution shifts when the flow speed is shocked to higher values, it is necessary to express the particle speed in $f_{\kappa}$ relative to flow speed in the spacecraft frame, that is, $v:\rightarrow v-V_{sw}^{\prime}$ \citep[see also][]{Leubner2004,Kongetal2017}.
The SW distribution is therefore expected to be transformed by the shock in at least three ways, depending on how $V_{sw}^{\prime}$, $n$ and $T$ change.
To illustrate this transformation, the SW distribution, as represented by $f_{\kappa}$ and viewed by an observer at Earth, is shown at different stages during the shock passage on the right-hand side of Figure \ref{fig:plasmakappas} for parameters as specified in Table \ref{tab:kappaparams}. 
Bear in mind these distributions have not yet been injected into the DSA process, and that these transformations occur solely because the shock had heated the SW plasma.
Also note that for consistency with the units in which observations are typically presented, $f_{\kappa}$ is converted throughout this study into units of differential intensity (as discussed in Appendix \ref{subsec:toward_j}) and is subsequently denoted as $j_{\kappa}$.

With the shock still 0.02 AU away from the Earth, Figure \ref{fig:plasmakappas} shows that the SW distribution at Earth is as of yet unchanged, since none of the plasma quantities have been shocked at this point. 
When the shock passes Earth, that is, when $r_{sh}=r_0$, the temperature is shown to have increased substantially, accompanied by more modest increases in the flow speed and density.
The most obvious changes the SW distribution incurred at this point is that it notably broadened as a result of the temperature increase, and that the thermal peak shifted to higher energies as a result of the increased flow speed.
Note that because $f_{\kappa}$ is normalised to the number density, the conservation of particles demands that the peak intensity of the distribution decreases when it broadens.
Considering, finally, how the distribution changes when the shock has moved 0.02 AU beyond the Earth, it shifts further, to higher energies, as the flow speed attains its full downstream value, while increasing overall intensities due to the increase in number density.
At this point the temperature had already plateaued and hence the distribution shows no appreciable broadening from when the shock passed Earth's position.

These SW distributions are specified as source spectra for DSA in Section \ref{subsec:sourcefunction}.

\section{Modelling energetic particle acceleration at CME shocks} \label{sec:ModelSDEs}

The events of interest in this study are large ESP events with small associated anisotropies. 
These events are typically associated with fast-moving shocks driven by halo CMEs \citep{Makelaetal2011}, that is, those propagating radially outward and approximately centred on the solar disk.
Therefore, to describe the transport and DSA of energetic particles associated with such events, it sufficient to solve the \cite{Parker1965} transport equation (TPE) for a single spatial dimension.
See also the motivation offered by \cite{Giacalone2015} in this regard.
Hence, in radial coordinates, the TPE is written as
\begin{align} \label{eqn:TPE} 
\frac{\partial f}{\partial t^{\prime}} &=  \left(\frac{1}{r^2}\frac{\partial}{\partial r}\left(r^2 \kappa_{rr}\right)-V_{sw}^{\prime}\right) \frac{\partial f}{\partial r} + \kappa_{rr}\frac{\partial^2 f}{\partial r^2} \nonumber \\ &+ \frac{1}{3r^2}
\frac{d \left(r^2 V_{sw}^{\prime}\right)}{dr}E\ \xi(E)\frac{\partial f}{\partial E} + Q \text{,} 
\end{align}
with $\xi(E)=(E+2E_p)/(E+E_p)$, where $E_p$ is the proton rest-mass energy. The TPE contains the relevant particle transport processes such as SW convection, spatial diffusion, and energy changes due to transport in regions of compressing or expanding SW flows.
The last-mentioned process implicitly simulates DSA for regions such as shocks with negative SW velocity divergences.
$Q$ represents a particle source function.
Note that the TPE is solved for the pitch-angle averaged, omni-directional distribution function $f=f_0 \left(r,p,t \right)$, which is only a function of position, scalar momentum and time; see Appendix \ref{sec:countingparticles}. 
As before, when presented, the distribution function is converted to units of differential intensity, that is, $j = p^2 f$.

The diffusion coefficient considered is the effective radial diffusion coefficient $\kappa_{rr}$, which relates to the equivalent mean free path (MFP) $\lambda_{rr}$ according to
\begin{equation} \label{eqn:difcoef}
\kappa_{rr} = \frac{v}{3}\lambda_{rr} \text{,}
\end{equation}
where $v$ is the particle speed and where $\lambda_{rr}$ is given by
\begin{equation}\label{eqn:MFPdef}
\lambda_{rr} = \lambda_0 \left(\frac{R}{R_0}\right)^{1/3}
\end{equation}
where $R = \sqrt{E^2+2EE_p}$ is the particle rigidity and where $\lambda_0$ is a reference MFP defined at $R_0 \equiv$ 1 GeV.
Given that the parallel diffusion coefficient is much larger than the perpendicular coefficient, and magnetic field lines are largely radial for the fast flow speeds considered, diffusion along field lines is assumed to dominate. 
The rigidity dependence implemented here is hence chosen to emulate that predicted for parallel diffusion by quasi-linear theory \citep{Jokipii1966} for a Kolmogorov-distributed turbulence power spectrum.
\cite{Kallenrodeetal1992} reports no pronounced variation in MFPs with radial distance between the Sun and 1 AU for diffusive events.
As such, a radial dependence is omitted in Eq. \ref{eqn:MFPdef}.
Furthermore, radial MFPs of 0.02 to 0.15 AU are reported for 0.3 to 0.8 MeV electrons \citep{Kallenrodeetal1992}, and the ratio of MFPs for 18 MeV protons to 1 MeV electrons as $1.6\pm 0.9$ \citep{Kallenrode1993}.
Taking the lower limits of the aforementioned quantities into account, the corresponding lower limit of $\lambda_0$ (defined at 1 GeV) for protons is estimated from Eq. \ref{eqn:MFPdef} as $\sim$0.05 AU.
$\lambda_0$ is varied in Section \ref{sec:spectral_features} and chosen as 0.06 AU elsewhere, using the above lower limit and the results of the aforementioned section as guidelines.
 
\subsection{The numerical model: SDEs} \label{subsec: SDEs}

The TPE specified in multiple computational dimensions typically requires numerical methods to solve. 
However, instead of solving Eq. \ref{eqn:TPE} using finite-difference methods \citep[e.g.][]{Giacalone2015}, the transport and DSA of energetic particles are simulated here by solving an equivalent set of SDEs \citep[see also][]{KruellsAchterberg1994,MarcowithKirk1999,Zhang2000}. 
The aspects of the SDE approach that are important for this study are detailed below. Refer to \cite{StraussEffenberger2017} for a comprehensive review.
 
Eq. \ref{eqn:TPE} is conveniently written in the form of the time-backward Kolmogorov equation %
\begin{align} \label{eqn:timeback_kolmgorov}
-\frac{\partial f}{\partial t} = \mu_r \frac{\partial f}{\partial r} + \mu_E \frac{\partial f}{\partial E} + \frac{1}{2} \sigma_r^2\ \frac{\partial^2 f}{\partial r^2} \text{,}
\end{align}
from which the SDEs can be cast into the form of the It{\^o} equation \citep[see e.g.][]{Zhang1999}
\begin{align} \label{eqn:genSDE_r}
dr &= \mu_r dt + \sigma_r dW \\
dE &= \mu_E dt \text{,} \label{eqn:genSDE_E}
\end{align}
where $dW = \sqrt{dt}\ \Lambda(t)$ represents the Wiener process and $\Lambda(t)$ is a simulated Gaussian-distributed pseudo-random number.
Note that the forward time $t^{\prime}$ as used in Eq. \ref{eqn:TPE} is related to the backward time $t$ according to $t=t_T-t^{\prime}$, where $t_T$ is the total simulation time. It thus follows that $\partial/\partial t^{\prime}=-\partial/\partial t$ in Eq. \ref{eqn:timeback_kolmgorov}.  
Substituting the coefficients corresponding to $\mu_r$, $\mu_E$, and $\sigma_r$ from Eq. \ref{eqn:TPE} into Eqs. \ref{eqn:genSDE_r} and \ref{eqn:genSDE_E} yields
\begin{align} \label{eqn:rad_SDE}
dr &= \left( \frac{1}{r^2}\frac{\partial}{\partial r}\left( r^2 \kappa_{rr} \right) - V_{sw}^{\prime} \right)dt + \sqrt{2 \kappa_{rr} dt}\ \Lambda(t)  \\
\label{eqn:energy_SDE}
dE &= \frac{1}{3r^2}
\frac{d \left(r^2 V_{sw}^{\prime}\right)}{dr} E\ \xi(E)\ dt  \text{,}
\end{align}
which are equivalent to the TPE in Eq. \ref{eqn:TPE}.

These SDEs are integrated in a time-backwards fashion: Starting from an observational point $(r_{\text{obs}},E_{\text{obs}})$ at which the value of the distribution function is sought, Eqs. \ref{eqn:rad_SDE} and \ref{eqn:energy_SDE} are solved iteratively using the Euler-Maruyama numerical scheme \citep{Maruyama1955} for a finite time step $\Delta t$. 
The coordinates $r$ and $E$ are updated upon each iteration until $t^{\prime}=0$, or equivalently, until $t=t_T$.
When simulating ESP events, $t_T$ is chosen as the transit time of the shock travelling (in a time-backwards fashion) from $r_{\text{obs}}$ to the inner modulation boundary near the Sun ($r_{\text{min}}=5R_{\odot}$), that is, $t_T=(r_{\text{obs}}-r_{\text{min}})/V_{sh}$.
The shock position $r_{sh}$ is therefore updated in step with $r$ and $E$, according to 
\begin{equation} \label{eqn:rshock}
r_{sh} = r_{\text{obs}}-V_{sh}t
\end{equation}
where $0<t\le t_T$.
A constant shock speed is assumed, since the mean acceleration of CME-driven shocks associated with ESP events is reportedly close to zero \citep{Makelaetal2011}. 
Note that the time is not incremented by the same amount during each iteration. 
It is instead specified to vary, depending on the dominant transport process at the current position, to maintain a fixed step length, that is,
\begin{equation} \label{eqn:dtvar}
\Delta t = \text{min}\left\lbrace 
	\dfrac{0.1L}{\text{max}\left(\left\lvert 
	\dfrac{1}{r^2}\dfrac{\partial}{\partial r}\left(r^2 \kappa_{rr} \right) 
	\right\rvert, V_{sw}^{\prime}\right)},\frac{0.1L^2}{\kappa_{rr}}  
	\right\rbrace \text{.}   
\end{equation}
This has at least two advantages: Firstly, limiting the step length to some fraction of the length scale $L$, which is associated with the shock, ensures that the shock structure is properly resolved. 
Secondly, scaling $\Delta t$ in this manner saves computation time, since instead of applying a very small fixed time step, $\Delta t$ will only be small when it is required.
As a result of this variable time step, computation times are typically longer for larger values of $\kappa_{rr}$, such as at higher energies, and for narrower shocks (smaller $L$-values). 

The above time-integration is repeated $N_p$ (typically, $\sim 10^{6}$) times for each observational point $(r_{\text{obs}},E_{\text{obs}})$, thereby tracing out $N_p$ trajectories in $r$ and $E$ of phase-space density elements, conventionally referred to as pseudo particles.
The source function $Q$ in Eq. \ref{eqn:TPE} is handled in the SDE approach as a correction term \citep{StraussEffenberger2017}. 
Expressing it as a rate of contribution to the distribution function allows the contribution per pseudo particle (or its \textit{amplitude}) to be calculated iteratively along the integration trajectory. 
That is, $Q = d\alpha/dt \implies \Delta\alpha = Q\Delta t$, from which it follows that 
\begin{equation} \label{eqn:particleweight}
\alpha_i(t-\Delta t) = \alpha_i(t) + Q\Delta t \text{.}
\end{equation}  
The source contribution is then tallied for $N_p$ pseudo particles such that
\begin{equation} \label{eqn:sourcecontribution}
f_\text{obs} \left( r_{\text{obs}},E_{\text{obs}} \right) = \frac{1}{N_p} \sum^{N_p}_{i=1} \alpha_i 
\end{equation}
gives the value of the distribution function at the observational point at time $t^{\prime}=t_T\  (t=0)$.
Section \ref{subsec:sourcefunction} discusses how the source function itself is specified.
Furthermore, physical particle distributions can also be obtained by tallying the amplitude-weighted flux contributions of pseudo particles within appropriately sectored spatial and energy intervals (or \textit{bins}) at any time during the simulation, divided by the number of pseudo particles counted within each sector. 
This method is used in Section \ref{subsec:accsites} to trace likely acceleration sites and seed-particle energies in a time-backwards fashion.

Finally, a reflective inner boundary is implemented such that if $r<r_{\text{min}} \implies r = 2r_{\text{min}}-r$, while at the outer boundary, if $r>r_{\text{max}}=1.4$ AU, the time-integration routine is interrupted and that pseudo-particle's contribution is discarded, defining an absorbing boundary condition.

\subsection{Modelling the shock source function} 
\label{subsec:sourcefunction}

\begin{figure}[tp]
\figurenum{3}
\centering
\includegraphics[scale=0.4]{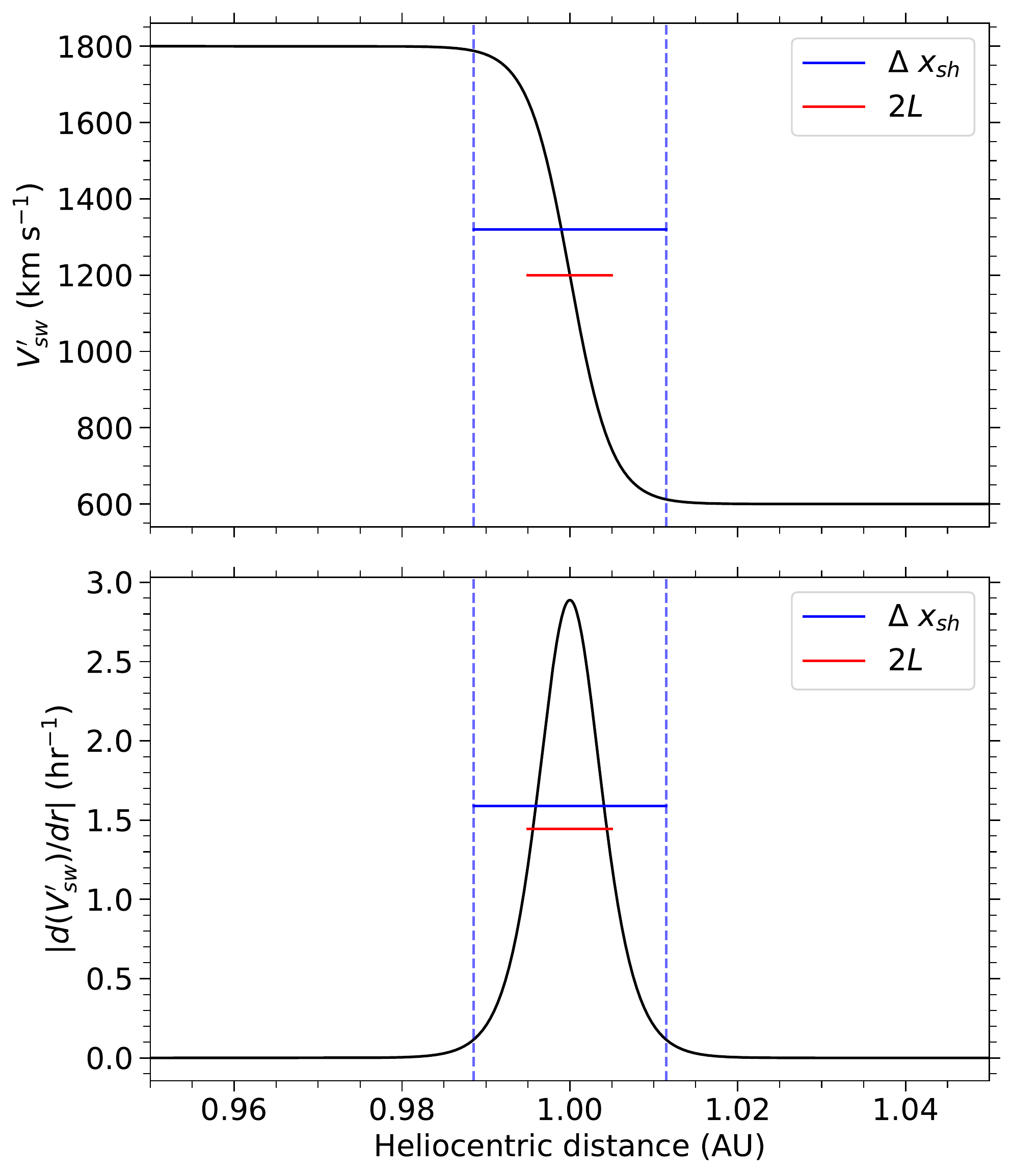}
\caption{Top: The transition of the fixed-frame SW flow speed $V_{sw}^{\prime}$ as function of heliocentric distance $r$ with the shock centred at Earth, that is, $r_{sh}=$ 1 AU. Bottom: Absolute value of the gradient of $V_{sw}^{\prime}$ corresponding to the profile shown above. $L\ (=0.005$ AU) and $\Delta x_{sh}$ are as defined for Eq. \ref{eqn:Vtransition} and Eq. \ref{eqn:L_relation}, respectively. \label{fig:source_width}}
\end{figure}

As per Eq. \ref{eqn:particleweight}, when a pseudo-particle is traced back to a point where it interacts with the shock, its amplitude is attributed a value that depends on the product of the source function $Q$ and the time $\Delta t$ it spends interacting with the shock.
Indeed, considering Eq. \ref{eqn:TPE}, $Q$ should have units corresponding to that of the distribution function per unit time.
Otherwise, the specification of $Q$ is largely arbitrary: e.g. \cite{Giacalone2015} employs delta functions, others \citep[e.g.][]{leRouxetal1996} specify distribution functions to represent seed populations, while \cite{MalkovVoelk1995} also consider shock-heated seed populations.
Drawing on aspects of the aforementioned examples, the following source function is proposed:
\begin{equation} \label{eqn:sourcefunction}
 Q = \frac{f_{\kappa}}{m_p^3}\ \left| \frac{dV_{sw}^{\prime}}{dr} \right| \ H\left(\frac{E-E_{\text{inj}}}{E_{\text{inj}}}\right) \text{.}
 \end{equation}
 Here, $f_{k}$ is the $\kappa$-function representing the SW distribution.
 As discussed in Appendix \ref{subsec:on_kappa_funcs}, the $m_p^3$ factor is included for dimensional consistency between $f_{\kappa}$ and the distribution function in the TPE.
 Since $f_{\kappa}$ is already normalised to the number density, scaling the SDE solutions is not necessary. 
The dimensionless Heaviside function $H\left((E-E_{\text{inj}})/E_{\text{inj}}\right)$ ensures that only the contributions of particles with energies larger than the injection energy $E_{\text{inj}}$ are included.
Note that $E_{\text{inj}}=$ 60 keV unless stated otherwise; see also Section \ref{subsec:Einjvar}.
The absolute value of the SW velocity gradient $\left| dV_{sw}^{\prime}/dr \right|$ is included to provide a spatial region with a width and position that is self-consistently associated with the shock and that has a reasonable probability of being sampled by pseudo particles.
Figure \ref{fig:source_width} shows how $\left| dV_{sw}^{\prime}/dr \right|$ is representative of the width of the shock. 
While it is sufficient for a physical particle to pass between the upstream and downstream media in order to gain energy, using this time-backward SDE method, a finite-width region has to be sampled by pseudo particles to register flux emanating from the shock.

\subsection{Advantages and limitations of the SDE approach}
\label{subsec:prosandcons}

The SDE approach has been successfully applied in many instances to simulate space particle transport \citep{Zhang1999,Peietal2010,Straussetal2011,Straussetal2013,Molotoetal2018} and DSA in particular \citep[][]{KruellsAchterberg1994,MarcowithKirk1999,Zhang2000,Zuoetal2011,Huetal2017}.
The time-backward approach is favoured here due to its efficiency, since every simulated pseudo-particle contributes to intensities at the desired observational point.
Additionally, in a similar fashion to how the boundary interactions of pseudo particles are used to trace the most probable points of entry into the heliosphere for cosmic rays \citep[e.g.][]{ Straussetal2011}, this backward tracing of phase-space trajectories is similarly utilised in this study to map probable acceleration sites and seed-particle energies; see Section \ref{subsec:accsites}.
Since these pseudo-particle trajectories are solved entirely independent of each other, this approach is also conducive to the utilisation of parallel computing platforms.

However, this mutual independence of the SDE solutions also limits applications to the test-particle case.
Non-linear effects of shock acceleration such as self-generated turbulence \citep{Lee1983,leRouxArthur2017} or particle mediation of shock structures \citep{Mostafavietal2017} can therefore not be considered.
Moreover, any process that requires the calculation of particle intensity gradients becomes computationally expensive \citep[e.g.][]{Molotoetal2018}.
Nevertheless, the SDE approach remains suitable to investigate the intricacies of classical DSA, without the limitations imposed by a numerical grid, such as instabilities involving the large gradients that are typically encountered at shocks. 
It is shown in Appendix \ref{sec:benchmark} that the SDEs produce appreciably similar results to a finite-difference approach in reproducing ESP events for the same parameter set.
Furthermore, as opposed to finite-differences schemes, the SDE model introduced in this study can seamlessly be expanded to a larger number of computational dimensions \citep[e.g.][]{Peietal2010} as required to study particle acceleration at more complicated shock geometries.  
  
\section{Spectral features and diffusion dependence of shock-accelerated particles.} \label{sec:spectral_features}

\begin{figure}[tp]
\figurenum{4}
\centering
\includegraphics[scale=0.4]{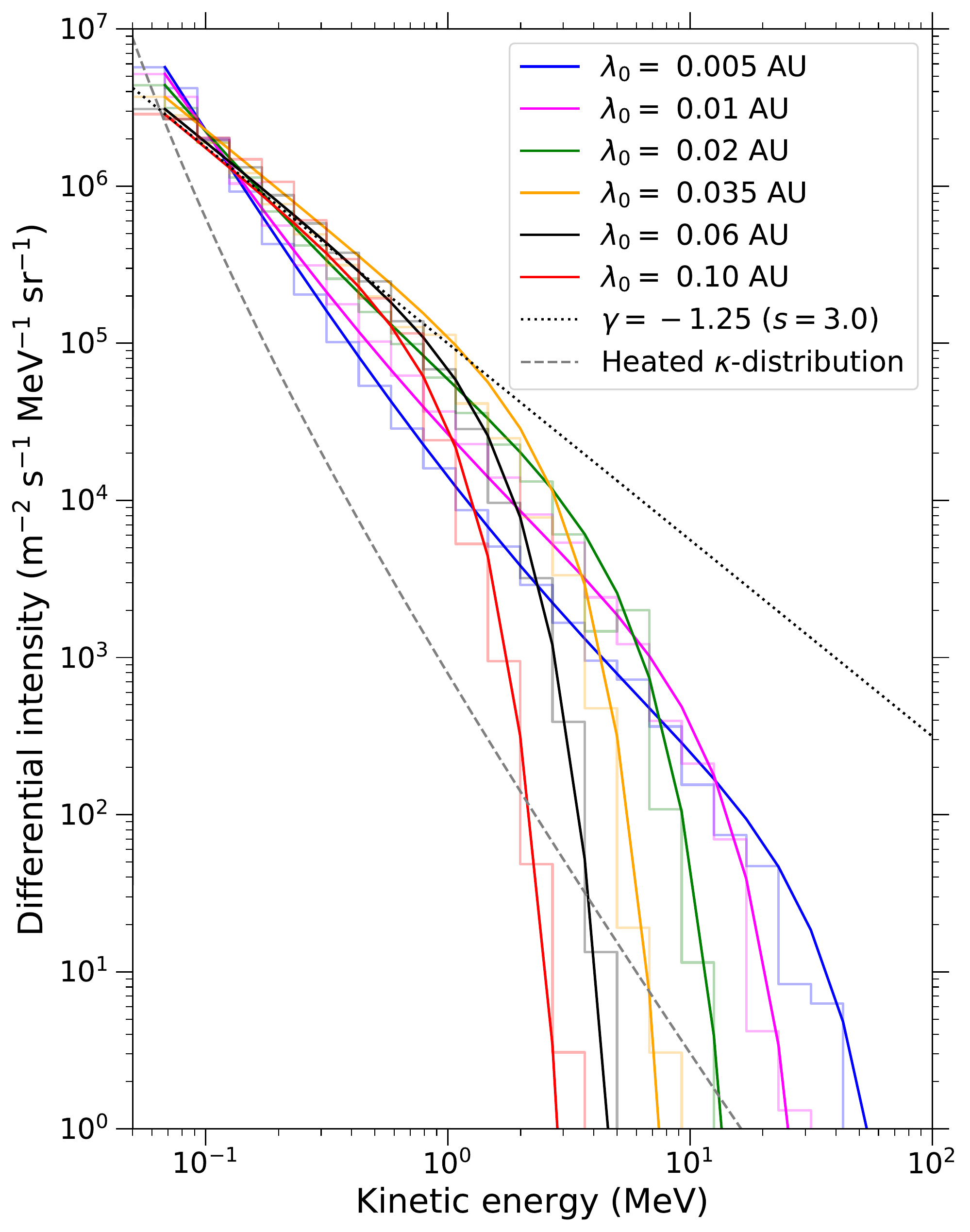}
\caption{Modelled energy spectra at the Earth ($r=$ 1 AU) at the time of the shock passage for different values of $\lambda_0$. Step-like lines represent SDE solutions, while the smooth solid lines are corresponding fits of Eq. \ref{eqn:genDSAform} using the parameters listed in Table \ref{tab:bestfit}. Also included are the heated $\kappa$-distribution (dashed grey line), specified as a source function on the shock, and the $E^{-1.25}$ power law associated with a shock compression ratio of $s=3$ (dotted line). \label{fig:spectra_lambdavar}}
\end{figure}

The most distinctive characteristics of shock-accelerated particles are observed in their energy spectra, which are considered here at the time the shock passes an observer, e.g. a spacecraft, near the Earth. 
Implementing the model configuration discussed in Sections \ref{sec:SWheating} and \ref{sec:ModelSDEs}, and scaling the diffusion coefficient by varying the value of $\lambda_0$ in Eq. \ref{eqn:MFPdef}, the dependence of shock-accelerated spectral features on this transport process is investigated.
The resultant spectra are shown in Figure \ref{fig:spectra_lambdavar}. 

Qualitatively, the typical DSA-associated features are evident: A power-law distribution at lower energies transitioning to an exponential-like decrease at higher energies \citep{EllisonRamaty1985}. This distribution can be described using a simple function in the form of
\begin{equation} \label{eqn:DSAform}
	j_{\text{DSA}} = j_0 \left(\frac{E}{E_0}\right)^{\gamma_s} \mathrm{e}^{ -\left(E/E_{\text{cut}} \right)^2}  \text{,}  
\end{equation}
 where $j_0$ is a differential intensity defined at some reference energy $E_0$, $E_{\text{cut}}$ is the cut-off (or roll-over) energy above which the distribution begins to decrease exponentially, and where 
\begin{equation} \label{eqn:specInd_DSA}
\gamma_s = \frac{s+2}{2-2s}
\end{equation} 
 is the spectral index (for $E \ll E_p$) associated with the compression ratio $s$ of the shock \citep[see also][]{Ellisonetal1990}. For $s=3$, as specified for these solutions, it is therefore expected that $\gamma_s=-1.25$.
However, not all of the spectra in Figure \ref{fig:spectra_lambdavar} appear to follow a $E^{-1.25}$ power law.
This is accentuated in Figure \ref{fig:specind_lambdavar}, which shows spectral indices for solutions with smaller $\lambda_0$-values actually varying with energy even before the exponential decreases ensue.
To describe this behaviour, it is useful to generalise Eq. \ref{eqn:DSAform} as follows:
  \begin{equation} \label{eqn:genDSAform}
	j_{\text{DSA}} = j_0 \left(\frac{E}{E_0}\right)^{\gamma_a} \left(\frac{E^{\xi}+E_{\text{tr}}^{\xi}}{E_0^{\xi}+E_{\text{tr}}^{\xi}} \right)^{\left(\gamma_b-\gamma_a\right)/\xi} \mathrm{e}^{ -\left(E/E_{\text{cut}} \right)^2} \text{.}  
\end{equation}
This describes a function transitioning between power-law indices $\gamma_a$ and $\gamma_b$ about $E_{\text{tr}}$, with $\xi$ specifying the smoothness of this transition, and which rolls over into an exponential decrease above $E_{\text{cut}}$.
Note that if $\gamma_a=\gamma_b$, this expression simplifies to the form of Eq. \ref{eqn:DSAform}.   
Setting $\gamma_b=\gamma_s$, where $\gamma_s=-1.25$ for $s=3$, and choosing $\xi=0.8$, the function given in Eq. \ref{eqn:genDSAform} is fitted to the solutions in Figure \ref{fig:spectra_lambdavar} with parameters as presented in Table \ref{tab:bestfit}.

\begin{table}
\centering
\begin{tabular}{rrrr}
\toprule
\multicolumn{1}{c}{$\lambda_0$ (AU)} & \multicolumn{1}{c}{$\gamma_a$} & \multicolumn{1}{c}{$E_{\text{tr}}$ (MeV)} & \multicolumn{1}{c}{$E_{\text{cut}}$ (MeV)}\\
\midrule
0.005 & $-2.5$ & 1.5 & 30.0 \\
0.01 & $-2.4$ & 0.5 & 11.0 \\
0.02 & $-2.0$ & 0.2 & 5.0 \\
0.035 & $-1.25$ & 0.08 & 2.5 \\
0.06 & $-1.25$ & - & 1.5 \\
0.1 & $-1.25$ & - & 0.9 \\
\bottomrule
\end{tabular}
\caption{Parameters used to fit Eq. \ref{eqn:genDSAform} to the corresponding model solutions shown in Figure \ref{fig:spectra_lambdavar}. The $\lambda_0$-values are constants used to scale the MFPs in Eq. \ref{eqn:MFPdef}. $E_{\text{tr}}$ and $E_{\text{cut}}$ are energies at which spectra transition to power-law indices associated with the full shock compression and roll over into exponential decreases, respectively. $\gamma_a$ is the spectral index for $E < E_{\text{tr}}$. Note that $E_{\text{tr}}$-values are omitted for the two largest $\lambda_0$-values, because $\gamma_a=\gamma_b=-1.25$, which reduces Eq. \ref{eqn:genDSAform} to the form of Eq. \ref{eqn:DSAform}.   }
\label{tab:bestfit}
\end{table}

Both $\gamma_a$ and $E_{\text{cut}}$ respond to varying the value of $\lambda_0$, affecting both the hardness of the spectra and the energies up to which they are accelerated. 
The behaviour of these two sets of features and the underlying physics are discussed separately in the subsections below. 

\subsection{Fractional compression sampling: On shock widths and diffusion length scales}
\label{subsec:fractional_compression}

\begin{figure}[tp]
\figurenum{5}
\centering
\includegraphics[scale=0.4]{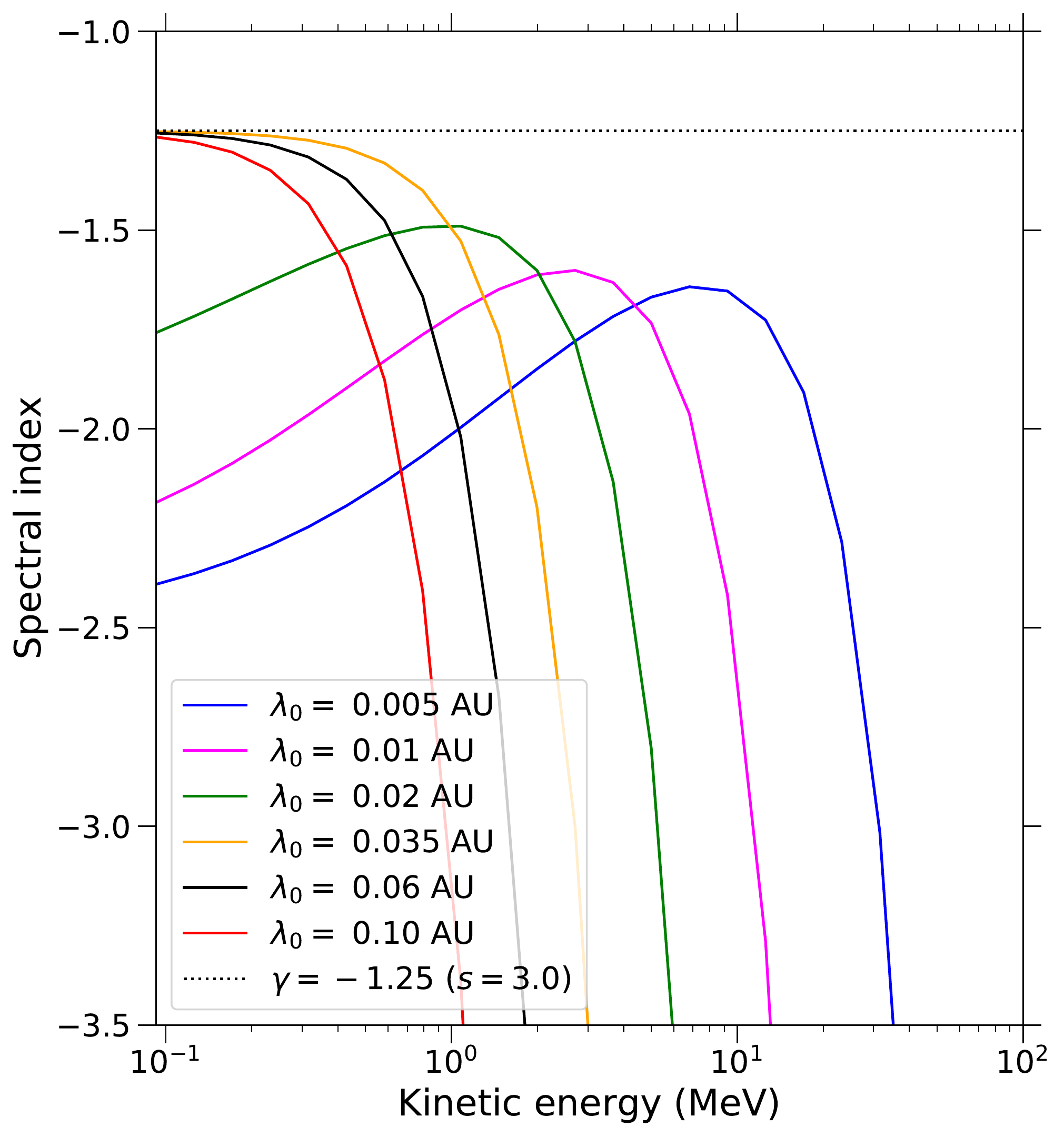}
\caption{Spectral indices for the corresponding fits of Eq. \ref{eqn:genDSAform} shown in Fig. \ref{fig:spectra_lambdavar}. The spectral index $\gamma_s=-1.25$ for a shock with $s=3$ is also shown. \label{fig:specind_lambdavar}}
\end{figure}

Classically, the spectral index associated with a DSA-produced spectrum is a function of the shock compression ratio alone.
The behaviour observed in Figure \ref{fig:spectra_lambdavar}, where shock-accelerated spectra become softer, displaying smaller spectral indices, for smaller diffusion coefficients, is therefore not theoretically expected.
Note, however, the solutions for $\lambda_0 \geq 0.035$ AU do follow the theoretically predicted power law of $E^{-1.25}$ for $s=3$.
Figure \ref{fig:specind_lambdavar} shows the spectral indices of the solutions presented in Figure \ref{fig:spectra_lambdavar}.
Here, spectral indices are also shown to be equal to $-1.25$ for large $\lambda_0$-values at low energies. 
Indices become progressively smaller for smaller $\lambda_0$-values, but increase toward higher energies.
Consider that particles with $\lambda_0 \lesssim 0.035$ AU may be sampling only a fraction of the total compression of the shock.
The spectra harden toward higher energies, because the MFPs themselves increase with energy and progressively greater fractions of the total compression are sampled. 

\begin{figure}[tp]
\figurenum{6}
\centering
\includegraphics[scale=0.4]{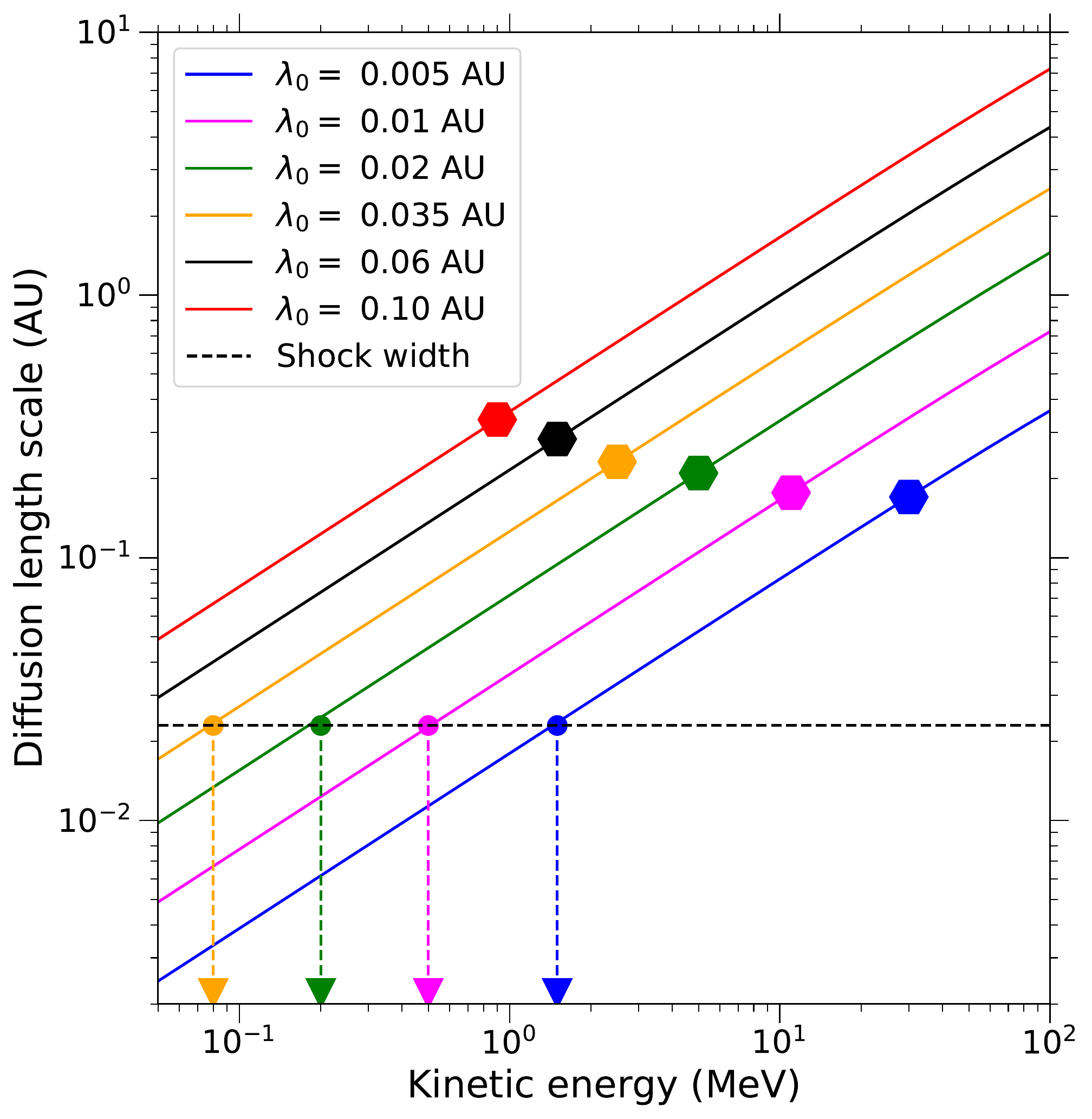}
\caption{Diffusion length scales at the shock, calculated using Eq. \ref{eqn:diflenscl} and corresponding to the solutions shown in Figure \ref{fig:spectra_lambdavar}. Varying $\lambda_0$-values scales the energy profiles of $\kappa_{rr}/V_{sw}$ uniformly. Markers indicate the values of $\kappa_{rr}/V_{sw}$ at the corresponding cut-off energies ($E_{\text{cut}}$) listed in Table \ref{tab:bestfit}. Arrows indicate the energies $E_{\text{tr}}$ (also listed in Table \ref{tab:bestfit}) at which the respective $\kappa_{rr}/V_{sw}$ profiles are equal to the shock width (dashed horizontal line), which is calculated using Eq. \ref{eqn:L_relation}. \label{fig:difscl_lambdavar}}
\end{figure}

This effect is illustrated in Figure \ref{fig:difscl_lambdavar}.
Recall from Eq. \ref{eqn:difcoef} that $\kappa_{rr} \propto \lambda_{rr}$, which is scaled using $\lambda_0$. 
Since $\kappa_{rr}$ does not have dimensions of length, when comparing to other length scales it is useful to define the diffusion length scale $\kappa_{rr}/V_{sw}$, which is often expressed as the sum of the down- and upstream values at the shock \citep[e.g.][]{SteenbergMoraal1999}. That is,
\begin{align} \label{eqn:diflenscl}
\left(\frac{\kappa_{rr}}{V_{sw}}\right)_{\left( r=r_{sh} \right)} &= \left(\frac{\kappa_{rr}}{V_{sw}}\right)_{1} + \left(\frac{\kappa_{rr}}{V_{sw}}\right)_{2}\nonumber\\  &= \kappa_{rr}\left( \frac{1}{V_{1}}+\frac{1}{V_{2}} \right) \text{,}
\end{align}   
where $(\kappa_{rr})_1=(\kappa_{rr})_2=\kappa_{rr}$, since $\kappa_{rr}$ is assumed not to change across the shock, with subscripts 1 and 2 denoting up- and downstream values, respectively. Here, $V_{1}=V_{sh}-V_1^{\prime}$ and $V_{2}=(V_{sh}-V_1^{\prime})/s$ are the up- and downstream flow speeds in the shock frame.
This quantity is plotted as a function of energy in Figure \ref{fig:difscl_lambdavar}, along with the shock width, calculated using Eq. \ref{eqn:L_relation} for $L = 0.005$ AU.
Note that for $\lambda_0=0.06$ and $0.1$ AU, for which spectra are aligned with $E^{-1.25}$, the diffusion length scales are greater than the shock width for all energies in the considered domain.
Those particles therefore sample the full compression ratio, and hence their energy distributions display the power law associated with $s=3$.
A similar effect is reported for the heliospheric termination shock \citep[e.g.][]{ArthurLeRoux2013}.
For $\lambda_0<0.06$ AU, particles sample fractional compression ratios up to the energies where their respective diffusion length scales begin to exceed the shock width.
Indeed, the energies at which $\kappa_{rr}/V_{sw}$ and the shock width intersect provide good estimates for $E_{\text{tr}}$ in Table \ref{tab:bestfit}.

Since the prediction of DSA given in Eq. \ref{eqn:specInd_DSA} is only observed where $\kappa_{rr}/V_{sw}$ is larger than the width of the shock, these results are in agreement with the length scale hierarchy of $\kappa_{rr}/V_{sw} \gg \lambda_{rr} \gg \Delta x_{sh}$ required for classical DSA to be valid \citep{BlandfordOstriker1978,JonesEllison1991}.
Given the $L$-dependence of the variable time step (Eq. \ref{eqn:dtvar}), to limit computation times, the shock width in this study is chosen much broader ($L=0.005$ AU) than typical IP shock widths \citep[$L\sim 10^{-6}$ AU; e.g.][]{Sapunovaetal2017}.
Nevertheless, the results of the model should remain valid as long as the aforementioned length scale hierarchy is observed.

\subsection{Finite-time acceleration and the termination of shock-accelerated spectra}
\label{subsec:acctime_cutoffs}

The second set of features considered entails the highest energies attained by shock-accelerated spectra before intensities begin to decrease exponentially.
Figure \ref{fig:spectra_lambdavar} illustrates that spectra for smaller $\lambda_0$-values are accelerated to higher energies before terminating.
These energies, represented by $E_{\text{cut}}$, are listed in Table \ref{tab:bestfit}.
Since they notably respond to varying $\lambda_0$, it seems reasonable to expect that this spectral transition also occurs due to $\kappa_{rr}/V_{sw}$ attaining some characteristic length.
Shock-accelerated spectra have previously been reported to roll over due to diffusion length scales becoming comparable to system sizes or some related length of the shock geometry \citep{EllisonRamaty1985,SteenbergMoraal1999}. 
Indeed, Figure \ref{fig:difscl_lambdavar} shows the diffusion length scales corresponding to each of the cut-off energies of the solutions in Figure \ref{fig:spectra_lambdavar} are reasonably similar, distributed between 0.1 and 0.4 AU.
However, these do not resemble any obvious length scale in either the physical system or within the numerical model. 
Simulations where the position of modulation boundaries were varied yielded negligible effects on energy spectra, suggesting that system size is not the limiting factor in this instance.

\begin{figure}[tp]
\figurenum{7}
\centering
\includegraphics[scale=0.4]{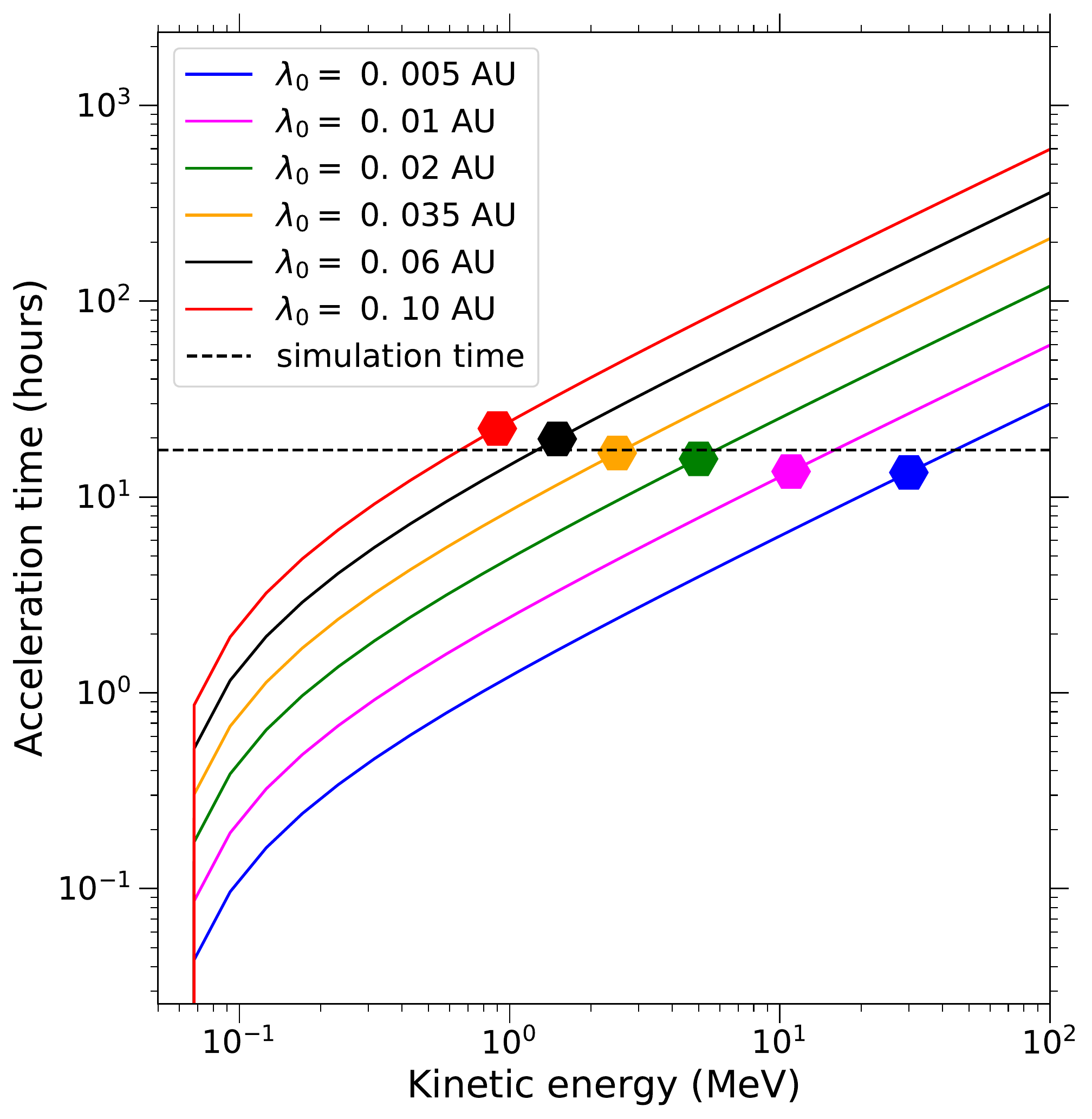}
\caption{Acceleration time as function of kinetic energy, calculated using Eq. \ref{eqn:acctime}, for the corresponding diffusion length scale profiles shown in Figure \ref{fig:difscl_lambdavar}. Markers indicate the time required to accelerate spectra up to the corresponding cut-off energies ($E_{\text{cut}}$) listed in Table \ref{tab:bestfit}. The simulation time (horizontal line) is the time taken by the shock to travel from near the Sun to Earth at 2400 km s$^{-1}$. \label{fig:acctime_lambdavar}}
\end{figure}

Another instructive quantity that is related to the diffusion length scale is the acceleration time \citep{Drury1983,Ellisonetal1990}, which is the time required to accelerate particles from a particular energy (or equivalent momentum) to another at a planar shock. This is expressed in terms of the diffusion length scale as 
\begin{equation} \label{eqn:acctime}
\tau_a = \frac{3}{V_{1}-V_{2}} \int\limits_{p_{\text{inj}}}^{p}\kappa_{rr}\left(\frac{1}{V_{1}}+\frac{1}{V_{2}}\right) \frac{dp^{\prime}}{p^{\prime}} \text{,}
\end{equation}
where $V_{1}$ and $V_{2}$ are as defined for Eq. \ref{eqn:diflenscl} and $p_{\text{inj}}$ is the momentum equivalent to $E_{\text{inj}}$. Note that momentum is generally related to kinetic energy according to $p=(1/c)\sqrt{E^2+2EE_{p}}$ (for protons), with $E_{p}$ the proton rest-mass energy.
Eq. \ref{eqn:acctime} is used in Figure \ref{fig:acctime_lambdavar} to approximate the time required to accelerate spectra up to different energies for each of the length scale profiles shown in Figure \ref{fig:difscl_lambdavar}.
Also shown in Figure \ref{fig:acctime_lambdavar} as symbols are the times required to accelerate spectra up to the observed cut-off energies listed in Table \ref{tab:bestfit} for each respective $\kappa_{rr}/V_{sw}$-profile.  
Similar to the diffusion length scales corresponding to the cut-off energies, the associated acceleration times also have similar values.  
These $\tau_a$-values also compare well to the total duration of the simulation, which is equal to the time taken by the shock to travel from near to the Sun to Earth at 2400 km s$^{-1}$, that is, $\sim$17.3 hours.  
This suggests that for each of the spectra presented in Figure \ref{fig:spectra_lambdavar}, $E_{\text{cut}}$ is the highest energy that could be attained in the available time \citep[see also][]{Channoketal2005}.
At $E > E_{\text{cut}}$, the rate of escape of particles from the shock begins to exceed the acceleration rate, which leads to the observed exponential intensity decreases.  
In this context, smaller $\lambda_0$-values, and consequently smaller diffusion coefficients, serve to better confine particles near the shock.
This, in turn, reduces the time required to accelerate particles up to a particular energy, or stated differently, allows particles to be accelerated to higher energies within the available time frame.

\subsection{Spectral features and acceleration efficiency}
\label{subsec:acc_efficiency}

\begin{figure}[tp]
\figurenum{8}
\centering
\includegraphics[scale=0.4]{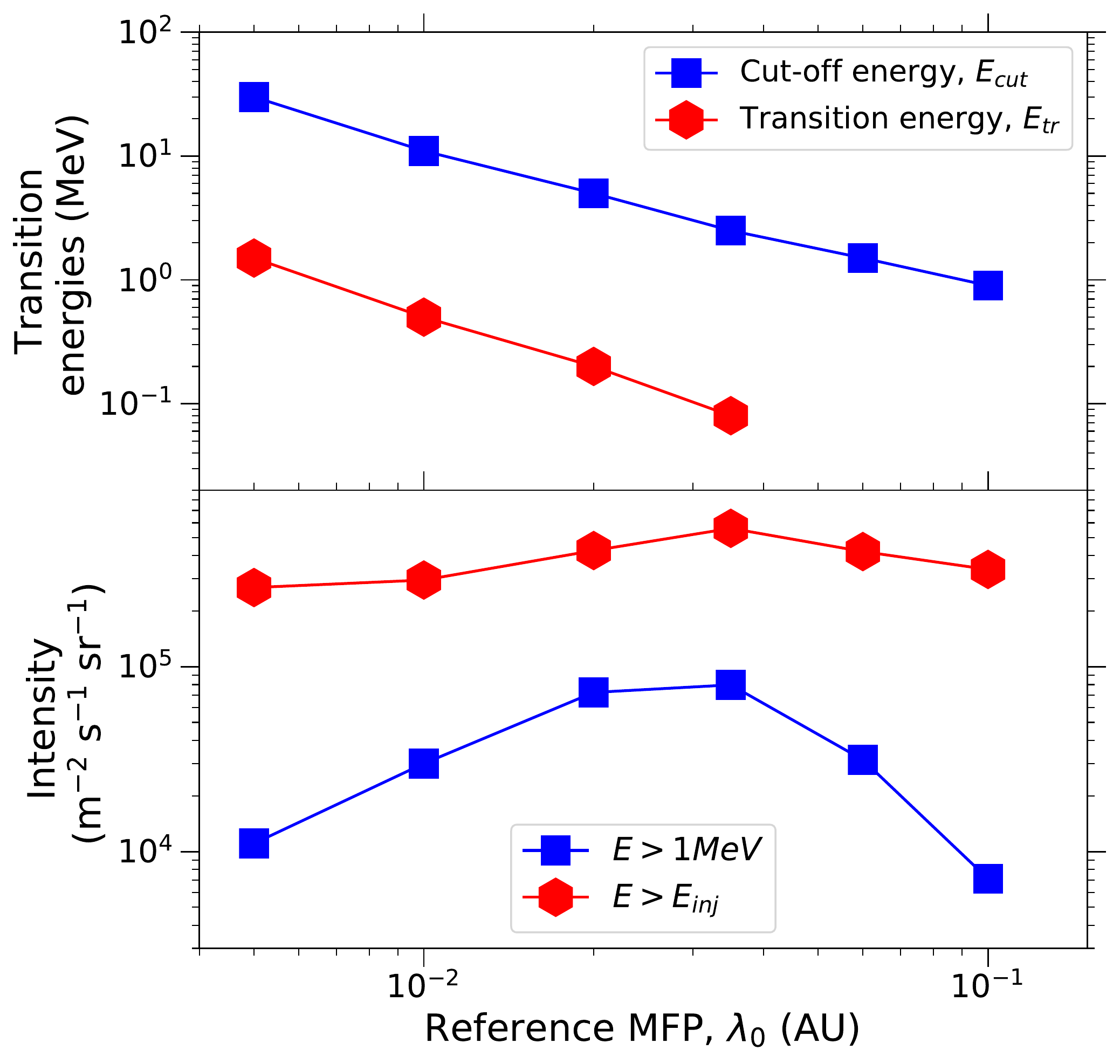}
\caption{Top: Energies $E_{\text{tr}}$ and $E_{\text{cut}}$ at which spectra for different values of $\lambda_0$ transition to power-law indices associated with the full compression and roll over into exponential decreases, respectively. Bottom: The energy-integrated distributions for different values of $\lambda_0$ as a measure of how efficiently DSA produces particles above the injection energy and 1 MeV, respectively. \label{fig:fluenceE_lambdavar}}
\end{figure}

The preceding discussions reveal that the characteristics of shock-accelerated spectra are sensitive to the value of diffusion length scales in two opposing ways: Diffusion length scales should exceed shock widths for spectra to display DSA-predicted power-law indices, but accelerated spectra terminate at lower energies for larger diffusion length scales.
The top panel of Figure \ref{fig:fluenceE_lambdavar} illustrates this dichotomy. 
The energies above which $\kappa_{rr}/V_{sw}$ exceeds the shock width, that is, $E_{\text{tr}}$, become smaller with increasing $\lambda_0$-values, implying greater overall spectral hardening, while similarly decreasing $E_{\text{cut}}$-values imply that spectra terminate at lower energies.   
Harder spectra yield larger intensities of energetic particles, but so do higher cut-off energies. 
Yet, with regards to diffusion properties, they are attained in opposite ways.
Neither of these therefore necessarily provide meaningful measures of acceleration efficiency on their own.

To evaluate which combination of spectral features yield greater intensities of energetic particles, it is useful to consider particle fluences. These are calculated by integrating differential intensities over energy as follows:
\begin{equation}\label{eqn:fluence}
I_{E_{l}} = \int\limits^{\infty}_{E_{l}} j(E)\ dE \text{.}
\end{equation}
This integral converges, since $j$ fortuitously always decreases as $E \rightarrow \infty$. 
Eq. \ref{eqn:fluence} is evaluated for each of the spectra shown in Figure \ref{fig:spectra_lambdavar} with two different lower limits, namely $E_{l} = E_{\text{inj}}$ and 1 MeV.
The resulting energy-integrated intensities are shown in the bottom panel of Figure \ref{fig:fluenceE_lambdavar}.   
The total intensity for all particles with $E>E_{\text{inj}}$ peaks at $\lambda_0\sim$ 0.035 AU, where the power-law segments of spectra have reached, or nearly reached, their maximum hardness.
$E_{\text{cut}}$ becomes more important when considering higher-energy particles.
For intensities integrated upward from 1 MeV, fluences peak between $\lambda_0= $ 0.02 and 0.035 AU and notably decrease for larger $\lambda_0$-values due to spectra rolling over at lower energies. 

It is revealed that both spectral hardness and the maximum energies spectra attain contribute appreciably to the number of energetic particles the DSA process is able to produce.
Of course, the energies up to which spectra extend become increasingly important for the intensities of higher-energy particles.

\section{Strong shocks versus fast shocks: Which is the more efficient particle accelerator?} \label{sec:strongvsfast}

Fast-moving CME shocks with large compression ratios are reported to be more efficient at accelerating energetic particles \citep{Larioetal2005b,Makelaetal2011,Giacalone2012}.
How conducive each of these shock properties are to producing large numbers of high-energy particles, especially as opposed to each other, raises an interesting subject for investigation.
Both the shock speed $V_{sh}$ and the compression ratio $s$ affect how particle distributions evolve, and often in contradicting (or even self-contradicting) ways.
Prior to injection, both properties affect the heating of the SW, while the hardness of shock-accelerated spectra is already shown to depend on the compression ratio.  
To explore and compare the effects of these two parameters, they are varied in three different ways: 
$V_{sh}$ and $s$ are each varied separately with the other remaining fixed, and they are varied together such that the factor by which the fixed-frame flow speed jumps across the shock remains constant.
For ease of reference, these three sets of parameter configurations are shown in Table \ref{tab:shspeedV}, along with the corresponding downstream flow speeds in both the fixed frame and the shock frame, denoted $V_{2}^{\prime}$ and $V_{2}$, respectively.

\begin{figure*}[tp]
\figurenum{9}
\centering
\includegraphics[scale=0.4]{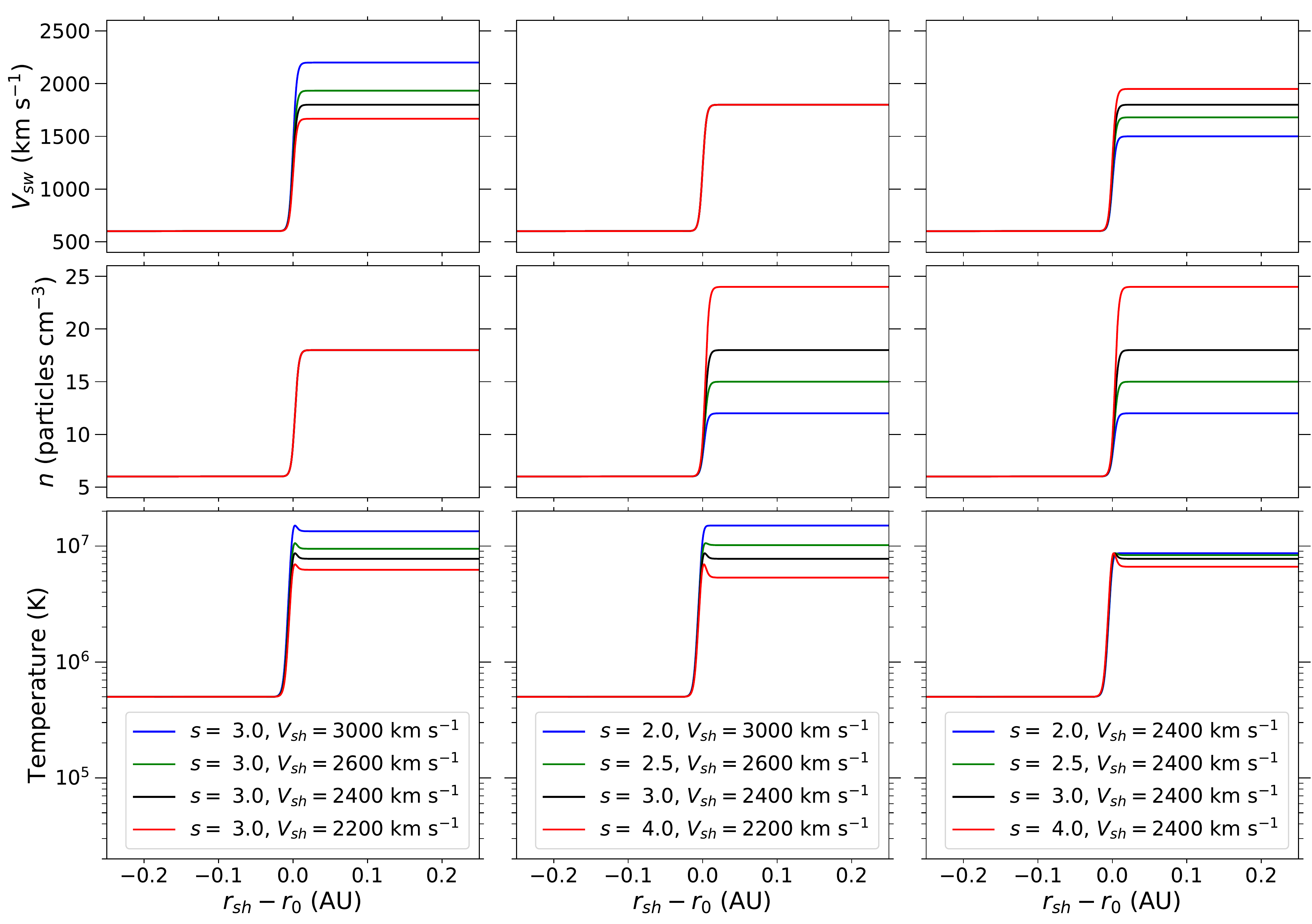}
\caption{Similar to the left-hand side of Figure \ref{fig:plasmakappas}, the profiles above are shown for some of the configurations of Table \ref{tab:shspeedV}, where the shock speed $V_{sh}$ and compression ratio $s$ are varied both separately, as shown in the left and right columns, respectively, and together as shown in the middle column. Note that profiles representing the reference configuration are shown in black.  \label{fig:plasma_shspeed}}
\end{figure*}

\begin{figure*}[tp]
\figurenum{10}
\centering
\includegraphics[scale=0.4]{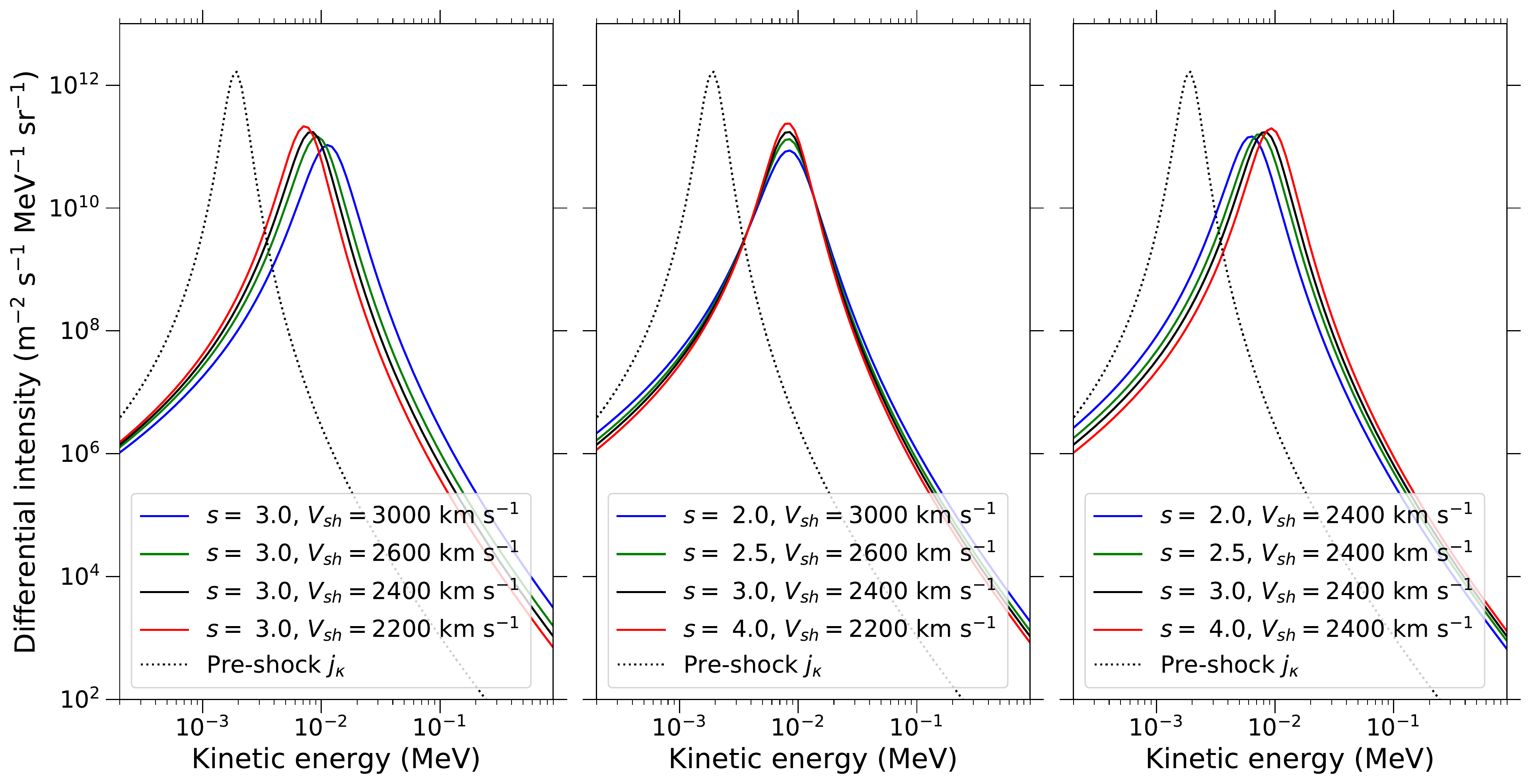}
\caption{The heating of the SW energy distribution at the Earth in response to the shock transitions of the plasma parameters shown in Figure \ref{fig:plasma_shspeed}. These distributions are shown for the same shock speed and compression ratio configurations used in the corresponding columns of Figure \ref{fig:plasma_shspeed}. The dotted lines represent the SW energy distributions from before the shock passage. \label{fig:kappas_shspeed}}
\end{figure*}

Varying $V_{sh}$ and $s$ affects the shock transitions of the SW flow speed, number density, and temperature, and thereby also influences how the energy distribution of SW particles changes during the passage of the shock.
Each of the aforementioned parameters' shock transitions are shown in Figure \ref{fig:plasma_shspeed} for the configurations listed in Table \ref{tab:shspeedV}.
The fixed-frame flow speed, that is, the flow speed as viewed by an observer stationed at Earth, increases with both $V_{sh}$ and $s$ when each is varied separately.
The results, as shown in the top row of frames in Figure \ref{fig:plasma_shspeed}, yield larger jumps in flow speed should either $V_{sh}$ or $s$ be increased.
As intended, when $V_{sh}$ and $s$ are varied together, the downstream flow speed remains constant in the fixed frame. 
Hence the factor by which the flow speed increases is the same for these instances.
The corresponding shock transitions of the number density and temperature in Figure \ref{fig:plasma_shspeed} follow from Eqs. \ref{eqn:ntransition} and \ref{eqn:dT_transition}.
Figure \ref{fig:kappas_shspeed} shows how the SW energy distributions change in response to these shock transitions.
It appears the SW energy distribution is heated by comparable amounts for varying shock speeds and strengths.
However, $j_{\kappa}$ appears marginally more sensitive to changes in temperature, which in turn is affected most appreciably by changes in $V_{sh}$.
Therefore, it can be argued that $V_{sh}$ contributes more towards heating the SW distribution, as opposed to $s$, considering the large shifts in energy and broadening of $j_{\kappa}$ that follows from varying it.

\begin{table}
\centering
\begin{tabular}{rrrrrr}
\toprule
& \multicolumn{1}{c}{$V_{sh}$} & \multicolumn{1}{c}{$s$} & \multicolumn{1}{c}{$V_{2}$ } & \multicolumn{1}{c}{$V_{2}^{\prime}$ } & \multicolumn{1}{c}{$V_{2}^{\prime}/V_1^{\prime}$}\\
& \multicolumn{1}{c}{(km s$^{-1}$)} & \multicolumn{1}{c}{} & \multicolumn{1}{c}{(km s$^{-1}$)} & \multicolumn{1}{c}{(km s$^{-1}$)} & \multicolumn{1}{c}{}\\
\midrule
1: & 3000 & 3 & 800 & 2000 & 3.67 \\
& 2800 & 3 & 733 & 2067 & 3.44 \\
& 2600 & 3 & 667 & 1933 & 3.22 \\
& 2280 & 3 & 560 & 1720 & 2.87 \\
& 2200 & 3 & 533 & 1667 & 2.78 \\
\midrule
2: & 2400 & 2 & 900 & 1500 & 2.5 \\
& 2400 & 2.2 & 818 & 1581 & 2.64 \\
& 2400 & 2.5 & 720 & 1680 & 2.8 \\
& 2400 & 3.5 & 514 & 1886 & 3.14 \\
& 2400 & 4 & 450 & 1950 & 3.25 \\
\midrule
3: & 3000 & 2 & 1200 & 1800 & 3 \\
& 2800 & 2.2 & 1000 & 1800 & 3 \\
& 2600 & 2.5 & 800 & 1800 & 3 \\
& 2280 & 3.5 & 480 & 1800 & 3 \\
& 2200 & 4 & 400 & 1800 & 3 \\
\midrule
ref:& 2400 & 3 & 600 & 1800 & 3.0 \\
\bottomrule
\end{tabular}
\caption{Configurations used throughout Section \ref{sec:strongvsfast}, in which 1: the shock speed $V_{sh}$ is varied, 2: the compression ratio $s$ is varied, and 3: $V_{sh}$ and $s$ are varied together such that $V_{2}^{\prime}/V_1^{\prime}$ remains constant. Here, $V_1^{\prime}=$ 600 km s$^{-1}$ and $V_{2}^{\prime}=(V_{sh}(s-1)+V_1^{\prime})/s$ are the fixed-frame flow speeds, respectively up- and downstream of the shock. $V_{2}=(V_{sh}-V_1^{\prime})/s$ is the downstream flow speed in the shock frame.
The last line contains the reference configuration.}
\label{tab:shspeedV}
\end{table}

\begin{figure*}[tp]
\figurenum{11}
\centering
\includegraphics[scale=0.4]{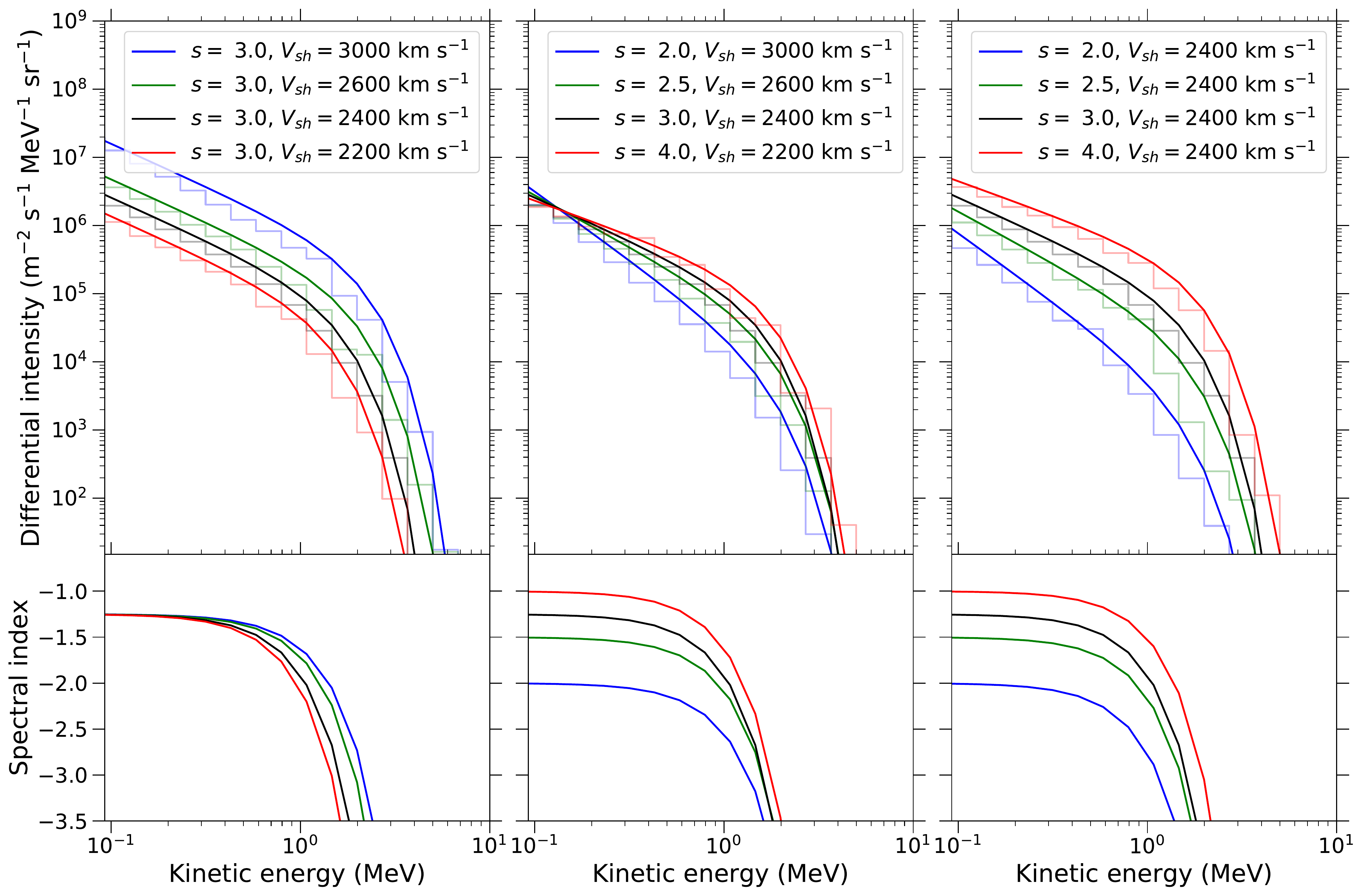}
\caption{Top: Modelled shock-accelerated energy spectra at the Earth at the time of the shock passage for some of the configurations listed in Table \ref{tab:bestfit_shspeedV} and with the corresponding SW distributions of Figure \ref{fig:kappas_shspeed} specified as input spectra. Step-like lines represent SDE solutions, while the smooth solid lines are corresponding fits of Eq. \ref{eqn:DSAform} using the parameters listed in Table \ref{tab:bestfit_shspeedV}. Bottom: Corresponding spectral indices for spectra shown above. \label{fig:spectra_Vshvar}}
\end{figure*}

The heated SW distributions discussed above essentially serve as input spectra, from which particles are injected into the DSA process at the shock for $E > E_{\text{inj}}$.
The shock-accelerated spectra obtained by solving the SDEs of Section \ref{subsec: SDEs} for each of the configurations in Table \ref{tab:shspeedV} are presented in Figure \ref{fig:spectra_Vshvar}. 
The analyses of these spectra are analogous to those of spectra presented in Figure \ref{fig:spectra_lambdavar}, while the discussions that follow draw on concepts introduced in Section \ref{sec:spectral_features}.
Accordingly, functions are fitted to the SDE solutions and the parameters tabulated for each configuration in Table \ref{tab:bestfit_shspeedV}.
The value of $\lambda_0$ is chosen as 0.06 AU throughout this section. 
With reference to Figure \ref{fig:difscl_lambdavar}, this implies the diffusion length scale exceeds the shock width for all considered energies and that the full compression ratio is sampled.
It is therefore sufficient to fit the single power-law function of Eq. \ref{eqn:DSAform}, with spectral indices as predicted by DSA.

The shock-accelerated spectra presented in Figure \ref{fig:spectra_Vshvar} share the same general features as those encountered before: a power-law segment extending from near the injection energy up to where it rolls over into an exponential decrease.
The following discussion considers how these features change as a result of varying the shock speed and compression ratio.
Firstly, varying $V_{sh}$, while keeping the compression ratio fixed at $s=3$, produces spectra as shown in the left column of Figure \ref{fig:spectra_Vshvar}. 
They are power-law distributed with a spectral index of $-1.25$, as illustrated in the accompanying frame below. 
This is the index expected from DSA for $s=3$ (see Eq. \ref{eqn:specInd_DSA}).
The sizeable differences in the intensities of these three spectra are due in large part to the heated SW spectra, which themselves are notably affected by varying $V_{sh}$.
When varying $s$ instead and keeping the shock speed fixed at $V_{sh}=$ 2400 km s$^{-1}$ as shown in the right-most column of Figure \ref{fig:spectra_Vshvar}, intensities differ mostly due to changes in the power-law indices of shock-accelerated spectra. 
These indices, shown in the accompanying frame below as $-2$, $-1.25$ and $-1$, correspond to those expected from Eq. \ref{eqn:specInd_DSA} for $s=2$, 3, and 4, respectively.

The middle column of Figure \ref{fig:spectra_Vshvar} illustrates the case where both $V_{sh}$ and $s$ are varied so that $V_{2}^{\prime}/V_1^{\prime}=3$. 
It is worth drawing attention to the fact that the spectral indices these spectra display are not the index associated with $V_{2}^{\prime}/V_1^{\prime}=3$, even though this would be the jump factor in the flow speed observed by spacecraft during the passage of the shock.
The power-law indices of all shock-accelerated spectra are consistently those associated with $s$.
The observed factor by which flow speeds change across a shock should not be confused with the actual compression ratio, which is the factor by which either the density or the flow speeds in the shock frame change across the shock.
This is important when calculating the power-law index of the DSA-predicted spectrum for comparison against an observed energy spectrum.

\begin{figure}[tp]
\figurenum{12}
\centering
\includegraphics[scale=0.4]{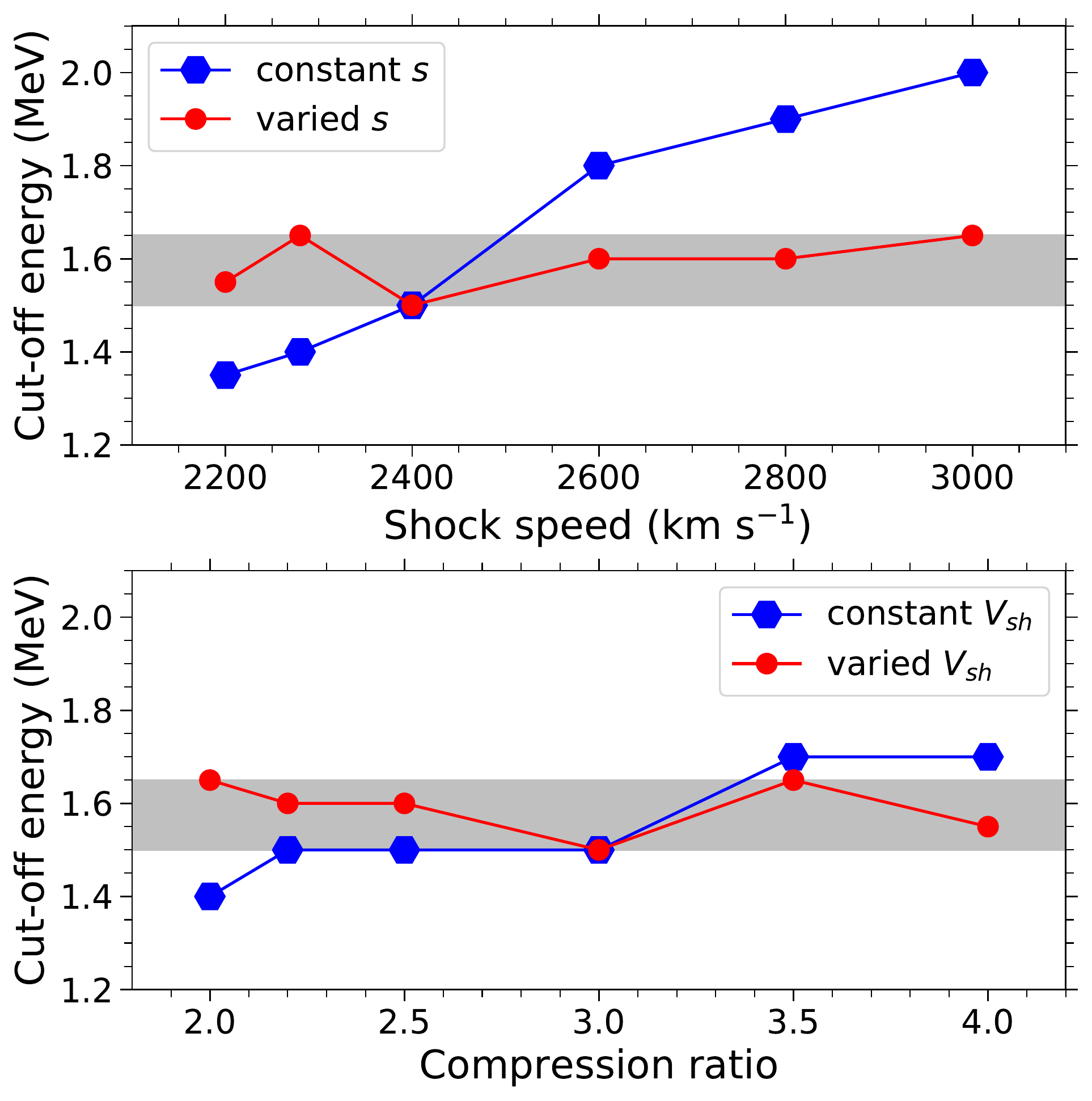}
\caption{Cut-off energies, where spectra roll over into exponential decreases, as functions of shock speed $V_{sh}$ and compression ratio $s$, for configurations where $V_{sh}$ and $s$ are varied both separately and together as specified in Table \ref{tab:bestfit_shspeedV}. \label{fig:Ecut_shspeed}}
\end{figure}

\begin{table}
\centering
\begin{tabular}{rrrrr}
\toprule
& \multicolumn{1}{c}{$V_{sh}$ (km s$^{-1}$)} & \multicolumn{1}{c}{$s$} & \multicolumn{1}{c}{$\gamma(s)$} & \multicolumn{1}{c}{$E_{\text{cut}}$ (MeV)}\\
\midrule
1: & 3000 & 3 & -1.25 & 2.0  \\
& 2800 & 3 & -1.25 & 1.9  \\
& 2600 & 3 & -1.25 & 1.8  \\
& 2280 & 3 & -1.25 & 1.4  \\
& 2200 & 3 & -1.25 & 1.35  \\
\midrule
2: & 2400 & 2 & -2.0 & 1.4  \\
& 2400 & 2.2 & -1.75 & 1.5  \\
& 2400 & 2.5 & -1.5 & 1.5  \\
& 2400 & 3.5 & -1.1 & 1.7  \\
& 2400 & 4 & -1.0 & 1.7  \\
\midrule
3: & 3000 & 2 & -2.0 & 1.65 \\
& 2800 & 2.2 & -1.75 & 1.6  \\
& 2600 & 2.5 & -1.5 & 1.6  \\
& 2280 & 3.5 & -1.1 & 1.65  \\
& 2200 & 4 & -1.0 & 1.55  \\
\midrule
ref:& 2400 & 3 & -1.25 & 1.5 \\
\bottomrule
\end{tabular}
\caption{Parameters used to fit Eq. \ref{eqn:DSAform} to the solutions of Figure \ref{fig:spectra_Vshvar} and corresponding to the configurations of shock speed $V_{sh}$ and compression ratio $s$ introduced in Table \ref{tab:shspeedV}.
Here, $\gamma(s)$ is the spectral index associated with $s$ as given by Eq. \ref{eqn:specInd_DSA} and $E_{\text{cut}}$ is the energy at which spectra roll over into exponential decreases.
The last line contains the reference configuration. }
\label{tab:bestfit_shspeedV}
\end{table} 

\begin{figure*}[tp]
\figurenum{13}
\centering
\includegraphics[scale=0.4]{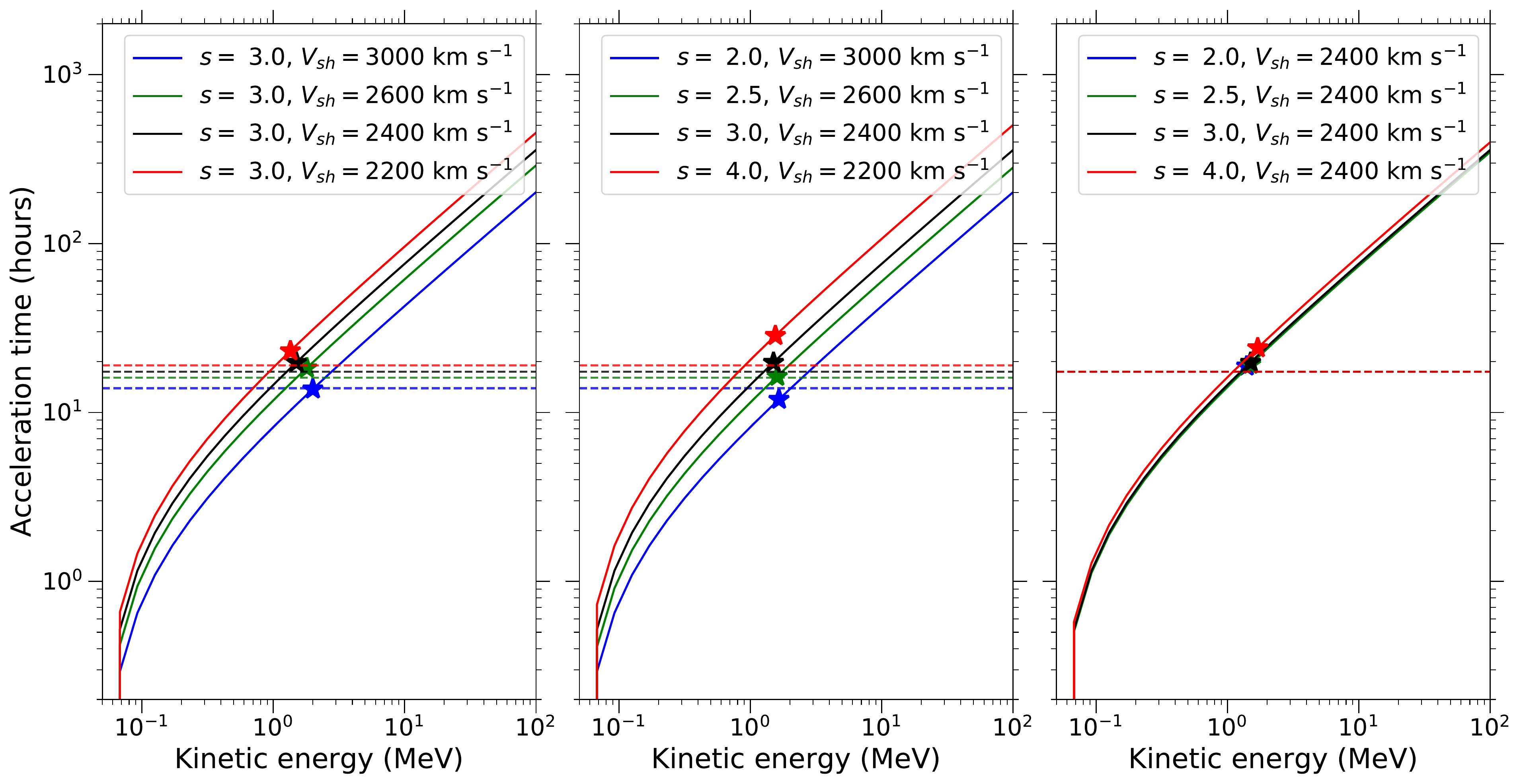}
\caption{Acceleration times similar to Figure \ref{fig:acctime_lambdavar} for each of the configurations of $V_{sh}$ and $s$ corresponding to the spectra shown in Figure  \ref{fig:spectra_Vshvar}. Markers indicate the time required to accelerate spectra up to the corresponding cut-off energies ($E_{\text{cut}}$) listed in Table \ref{tab:bestfit_shspeedV}. The total simulation times (indicated as dashed horizontal lines) correspond to configurations represented in the same color and indicate the travel times of the shock moving at $V_{sh}$ between the Sun and the Earth.  \label{fig:acctime_shspeed}}
\end{figure*}

The spectra shown in the left and right-hand sides of Figure \ref{fig:spectra_Vshvar} extend to higher energies for both faster and stronger shocks, where $V_{sh}$ and $s$ are each varied separately with the other held constant.
Spectra extend up to similar energies where $V_{sh}$ and $s$ are varied together.
This is also illustrated in Figure \ref{fig:Ecut_shspeed}, where the cut-off energies of the aforementioned spectra are plotted against both $V_{sh}$ and $s$, showing almost no reponse to changing shock conditions. 
Similar to Section \ref{subsec:acctime_cutoffs}, the acceleration times corresponding to each of the spectra in Figure  \ref{fig:spectra_Vshvar} are calculated and compared to the simulation times, as shown in Figure \ref{fig:acctime_shspeed}.
As before, the maximum attainable energies for shock-accelerated spectra are determined predominantly by the shock transit time.

\subsection{Acceleration efficiency: Strong versus fast shocks}
\label{subsec:acc_efficiency_svsv}

\begin{figure*}[hpt]
\figurenum{14}
\centering
\includegraphics[scale=0.4]{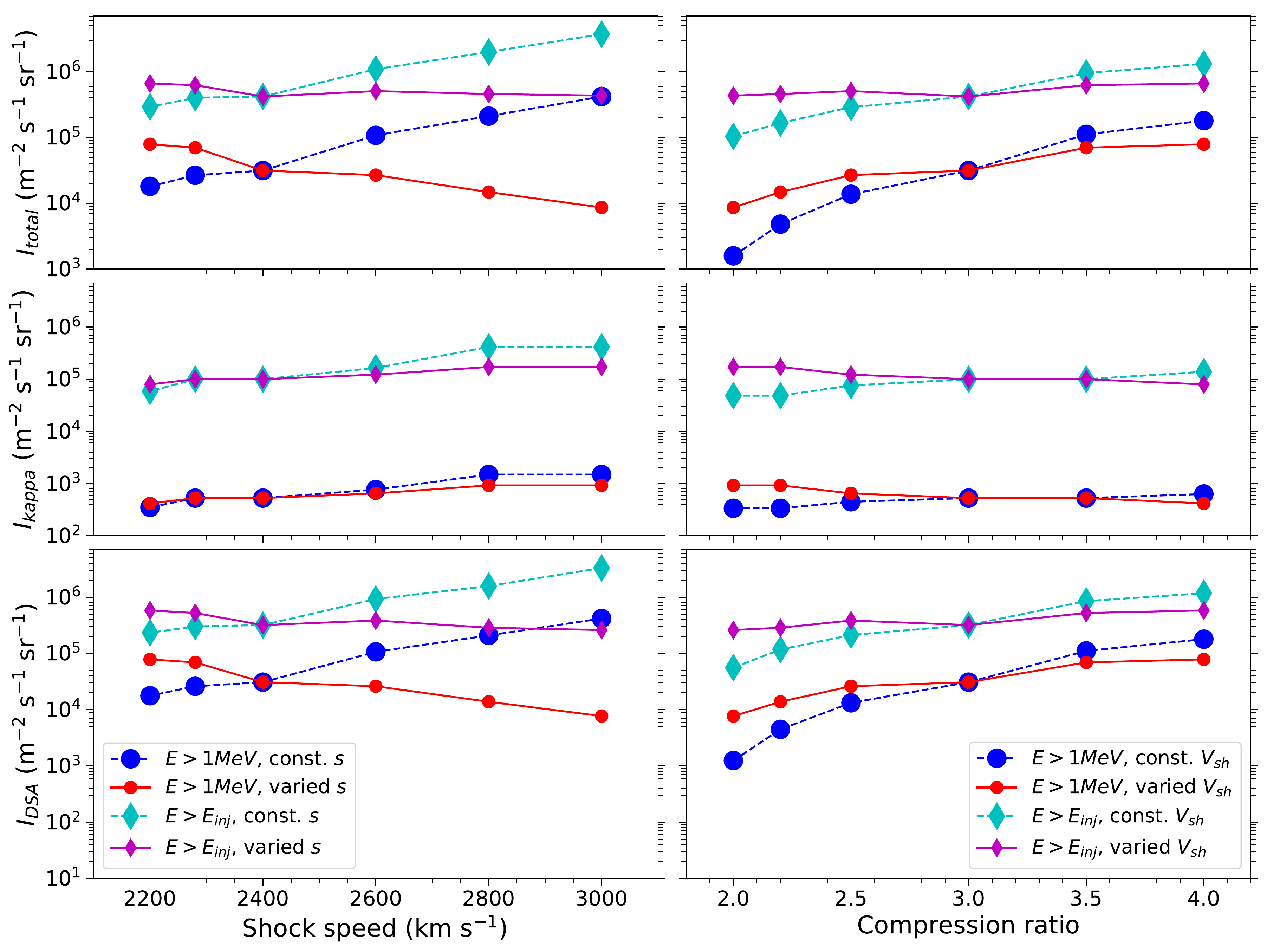}
\caption{The energy-integrated intensities (or fluences) of energetic particles above the injection energy ($E_{\text{inj}}$) and 1 MeV, represented by diamond and  dot markers, respectively, as functions of shock speed $V_{sh}$ and compression ratio $s$. $I_{\text{total}}$, $I_{\text{kappa}}$, and $I_{\text{DSA}}$ represent fluences for shock-accelerated spectra, heated SW distributions and those resulting from DSA alone, where $I_{\text{DSA}} = I_{\text{total}}-I_{\text{kappa}}$. $V_{sh}$ and $s$ are varied both separately and together for configurations as specified in Table \ref{tab:shspeedV}.  \label{fig:fluences_shspeed}}
\end{figure*}

The preceding sections analyse the collated effects of varying $V_{sh}$ and $s$ during the complete acceleration process, including the heating of the SW energy distribution and the subsequent shock acceleration.
It can be argued that the shock speed has the largest effect, since varying $V_{sh}$ yields pronounced changes in both the heated SW distributions and in reducing acceleration times, thereby extending spectra to higher energies.
On the other hand, the most notable effect of varying $s$ is the changes in the power-law indices displayed by shock-accelerated spectra, which also affects energetic particle intensities considerably.

Implementing the same technique used in Section \ref{subsec:acc_efficiency}, particle fluences are calculated using Eq. \ref{eqn:fluence}.
The resulting energy-integrated intensities are shown in the top frames of Figure \ref{fig:fluences_shspeed} as functions of both $V_{sh}$ and $s$.
To constrain the intensities resulting solely as a result of DSA and not including those of the input spectra, fluences are also calculated for the heated SW distributions ($I_{\text{kappa}}$)  and subtracted from those calculated for the shock-accelerated spectra ($I_{\text{total}}$).      
These two additional sets of energy-integrated intensities, namely $I_{\text{kappa}}$ and $I_{\text{DSA}}=I_{\text{total}}-I_{\text{kappa}}$, are also shown in Figure \ref{fig:fluences_shspeed}.
Note that $I_{\text{kappa}}$ is roughly an order of magnitude smaller than $I_{\text{total}}$ on average, with comparably weak dependences on $V_{sh}$ and $s$.
$I_{\text{DSA}}$ therefore closely resembles $I_{\text{total}}$.
Considering particle fluences as functions of $V_{sh}$ and $s$, where each is varied alone, it is apparent that both faster and stronger shocks yield greater numbers of energetic particles, whether considered for $E>E_{inj}$ or for $E>$ 1 MeV.
Since compression ratios for shocks in the SW do not typically exceed 4, and there is no hard limit on the speed shocks can attain, faster shocks arguably have greater potential as particle accelerators.

It is insightful to consider fluences for the configurations where $V_{sh}$ and $s$ are varied together, since this is essentially comparing the acceleration efficiencies of strong slower-moving shocks against weaker fast-moving shocks.
For these cases, fluences remain fairly constant for all combinations of $V_{sh}$ and $s$ for $E>E_{\text{inj}}$.
However, if only the number of higher-energy particles is considered, that is, for $E >$ 1 MeV, the results are more interesting:
As a function of compression ratio, the fluences increase as shocks become stronger, despite simultaneously becoming slower.
 As a function of the shock speed, these fluences become smaller for faster shocks for which the accompanying compression ratios are smaller.
This implies that strong slower-moving shocks are able to produce a greater number of energetic particles than weak fast-moving shocks, at least within the parameter ranges considered and for shocks with $V_2^{\prime}/V_1^{\prime}=3$.

\section{From the solar wind to energetic storm particles} \label{sec:SWtoESPs}

\begin{figure}[tp]
\figurenum{15}
\centering
\includegraphics[scale=0.4]{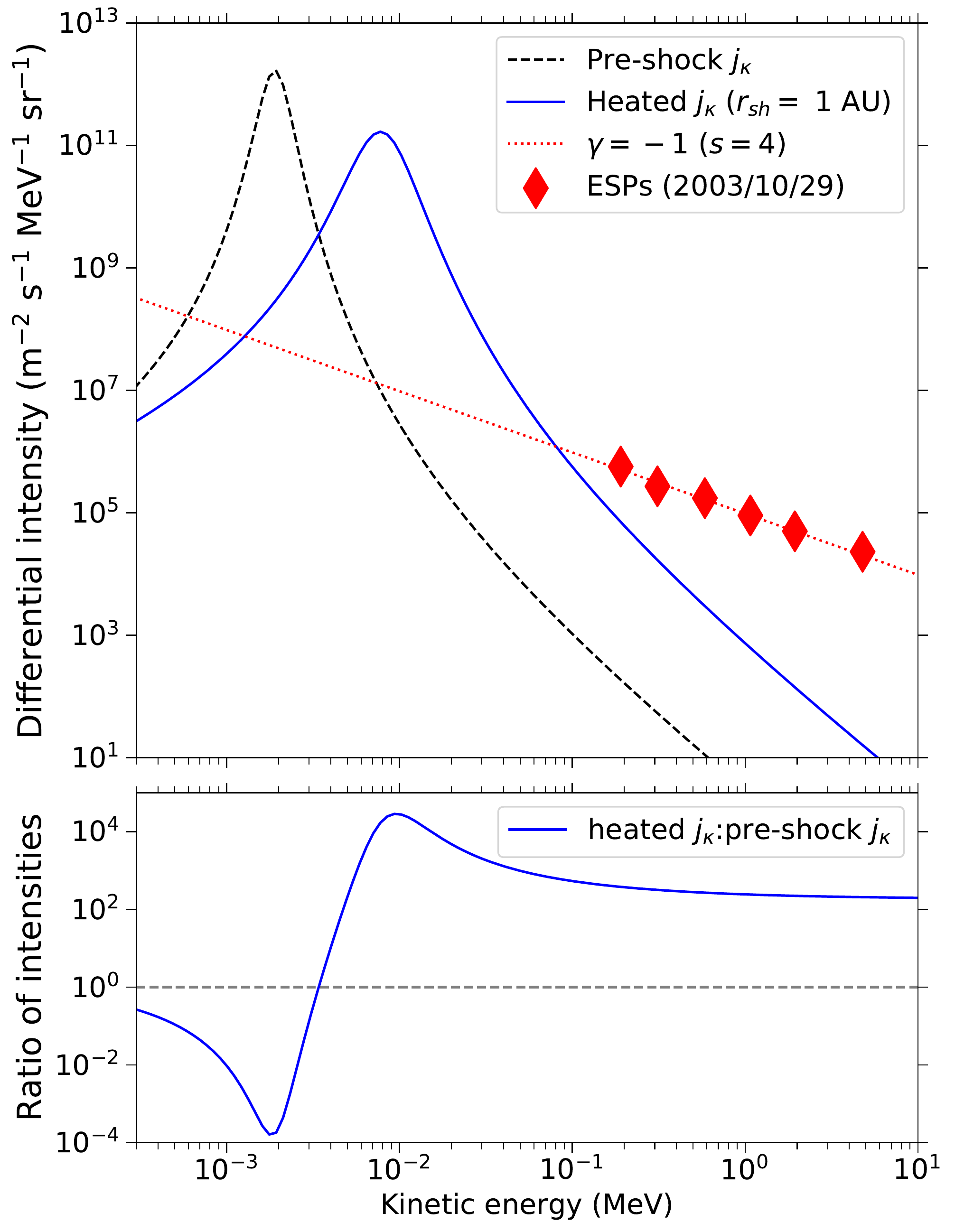}
\caption{Top: The undisturbed and heated SW energy distributions at Earth, with the latter corresponding to the time at which the shock passes the Earth's position. For reference purposes, the intensities observed by \textit{ACE}/EPAM LEMS30/120 during an energetic particle event on 2003 October 29 is also shown. These observed intensities can be fitted with the $E^{-1}$ power law associated with a strong shock of $s=4$. Bottom: Intensity ratios of the heated and undisturbed SW distributions shown above.  \label{fig:heatingratio}}
\end{figure}

In this section, the development of particles injected from the suprathermal SW into energetic particles associated with ESP events is considered in greater detail.
The underlying question of how SW particles can be accelerated to the much higher energies at which ESPs are observed is visually represented in Figure \ref{fig:heatingratio}.
The intensities observed during the \textit{Halloween} ESP event of 2003 October 29 are shown as an example of typical energy spectra observed during such events.
Note this particular spectrum is reproduced well by the power law associated with $s=4$ \citep[e.g.][]{Giacalone2015}.
  
The acceleration process is initiated with the heating of the SW energy distribution, where the $\kappa$-function describing it is allowed to change in response to changes in flow speed, density and temperature across the shock as shown in Section \ref{sec:SWheating}.
The importance of this initial heating cannot be understated:
consider, firstly, a spectrum shock-accelerated from the original, undisturbed SW distribution.
It would likely not extend up to the energies of observed ESPs because of the limited time available for acceleration.
Secondly, assuming the injection threshold for DSA is much larger than thermal energies, the larger intensities associated with the heated SW could allow shock-accelerated intensities to reproduce those of typical ESPs.  
Indeed, the bottom panel of Figure \ref{fig:heatingratio} illustrates that the heated SW distribution increases intensities of potential DSA seed particles by up to a factor of $10^{4}$, while increasing suprathermal tail intensities a hundredfold at least.
Interestingly, observations suggest that peak fluxes during ESP events display some dependence on CME sheath temperature \citep{Dayehetal2018}, while the broadening of SW distributions (which influence eventual intensities of shock-accelerated particles) is predominantly dictated by temperature changes across the shock.

\begin{table}
\centering
\begin{tabular}{rrrr}
\toprule
\multicolumn{1}{c}{$E_{\text{inj}}$ (keV)} & \multicolumn{1}{c}{$\gamma_a$} & \multicolumn{1}{c}{$E_{\text{tr}}$ (keV)} & \multicolumn{1}{c}{$E_{\text{cut}}$ (MeV)}\\
\midrule
10 & $-2.0$ & 30 & 0.7 \\
30 & $-1.8$ & 30 & 1.0 \\
60 & $-1.25$ & - & 1.5 \\
\bottomrule
\end{tabular}
\caption{Parameters used to fit Eq. \ref{eqn:DSAform} to the solutions of Figure \ref{fig:spectra_einjvar} for different injection energies $E_{\text{inj}}$. All quantities are as described for Table \ref{tab:bestfit}. }
\label{tab:bestfit_Einj}
\end{table}  

\subsection{Injection from the heated SW distribution} \label{subsec:Einjvar}

In the simplest terms, the injection threshold for DSA can be considered the energy at which particles should at least propagate upwind in order to make repeated shock crossings. 
This injection energy can be inferred by matching theoretically expected shock-accelerated spectra with observations \citep[e.g.][]{NeergaardParkerZank2012}.
It can also be estimated using the argument that the particle anisotropy must be small in order for DSA to be valid \citep{GiacaloneJokipii1999,Zanketal2006}.
For the typical values implemented in this study ($V_1^{\prime}=$ 600 km s$^{-1}$, $s=3$), injection energies are in the order of a few keV for parallel shocks, for which injection from an unheated Maxwellian seed particle distribution would indeed be possible \citep[][]{NeergaardParkerZank2012}.
However, for quasi-perpendicular shocks, and in the absence of magnetic field-line wandering, it is estimated that $E_{\text{inj}}\gtrsim$ 0.1 MeV \citep[see also][]{Li2017}.
The prior heating of the SW distribution becomes especially useful in this situation.
See also the simplified method used by \cite{Huetal2017} to estimate injection energies.   
  
\begin{figure}[tp]
\figurenum{16}
\centering
\includegraphics[scale=0.4]{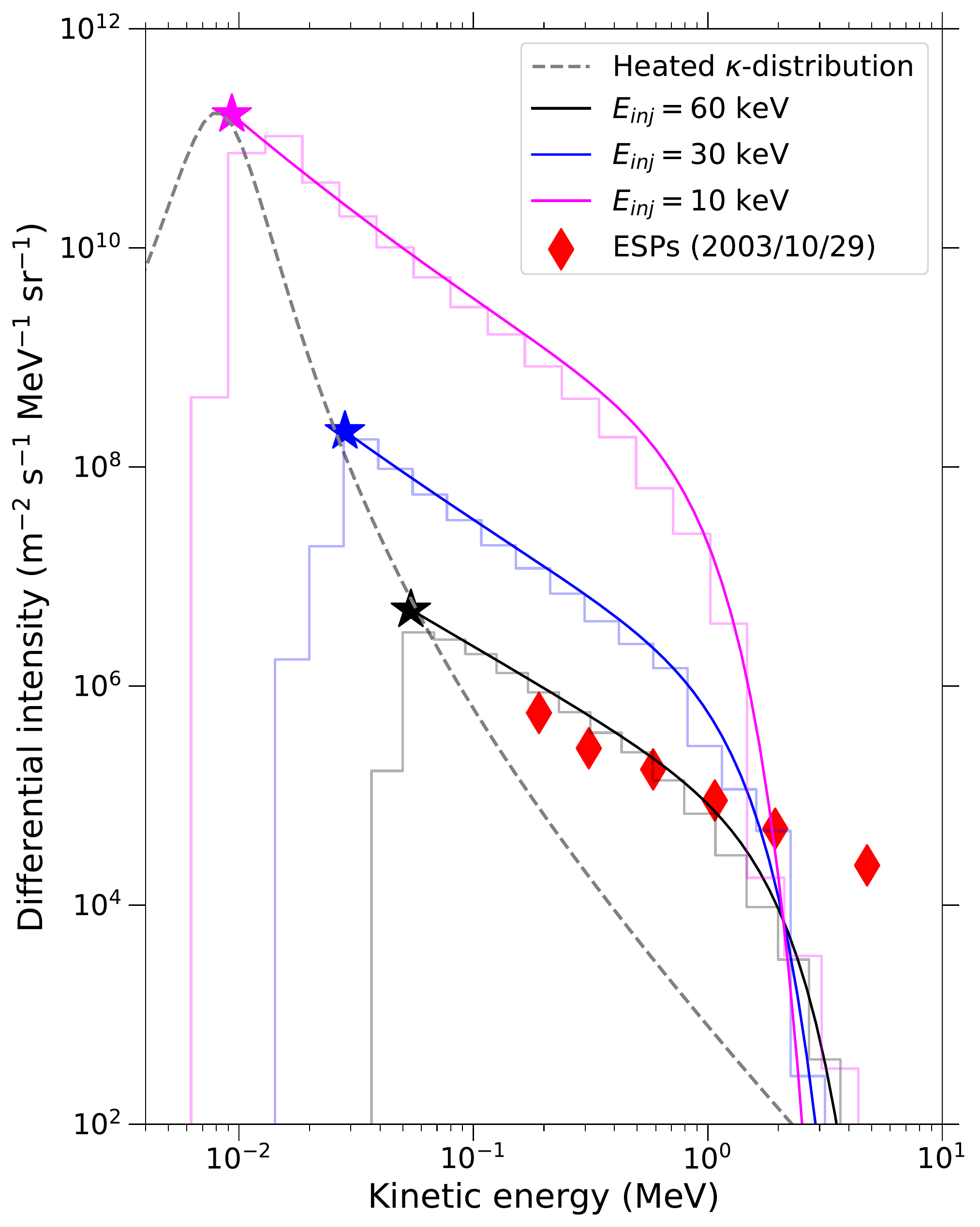}
\caption{Modelled shock-accelerated spectra at the Earth at the time of the shock passage for different injection energies. Step-like lines represent SDE solutions while the solid lines are fits of Eq. \ref{eqn:genDSAform} with parameters as listed in Table \ref{tab:bestfit_Einj}. The dashed grey line represents the heated SW distribution. The ESP observations are the same as shown in Figure \ref{fig:heatingratio}. \label{fig:spectra_einjvar}}
\end{figure}  

The halo-CME shocks considered in this study are mostly quasi-parallel if a \cite{Parker1958} spiral is assumed for the magnetic field.    
Here, $E_{\text{inj}}$ is varied between 10 and 60 keV, spanning an energy domain extending from the thermal peak to the suprathermal tail of the heated SW distribution, and bounded by the aforementioned estimates for parallel and quasi-perpendicular shocks.
The effects of these varied injection thresholds on shock-accelerated spectra are shown in Figure \ref{fig:spectra_einjvar}.
As before, Eq. \ref{eqn:genDSAform} is fitted for each case and the parameters listed in Table \ref{tab:bestfit_Einj}. 
The reference configuration as specified in Table \ref{tab:shspeedV} is used, with $\lambda_0=$ 0.06 AU, $V_{sh}=$ 2400 km s$^{-1}$, and $s=3$.
The standard features are visible: the spectra are distributed according to $E^{-1.25}$ as expected for $s=3$ and rolls over exponentially at higher energies.
Spectra appear to terminate at similar energies for the injection thresholds considered, as this is presumably governed by the acceleration time.

It can be seen from Figure \ref{fig:spectra_einjvar} that injecting particles from the heated SW distribution yields high intensities in the energy domain typically associated with ESPs.
In the case of $E_{\text{inj}}=$ 60 keV, the shock-accelerated intensities are quite similar to those observed for the ESP event included in Figure \ref{fig:spectra_einjvar} as an example.
Should the injection speed be defined as the minimum possible speed a particle needs to stay ahead of the shock, whilst following a Parker field line with a 45$^{\circ}$ angle ahead of the shock nose, the injection speed can be written from focused transport theory as $v_{\text{inj}}=\lvert V_1^{\prime} - V_{sh} \rvert/ \cos\left(45^{\circ}\right)$.
For a fast halo-CME shock with $V_{sh}= 2400$ km s$^{-1}$ and a more typical upstream flow speed of 400 km s$^{-1}$, the injection speed at 1 AU in the upstream flow frame is $v_{\text{inj}}=$ 2828 km s$^{-1}$, which corresponds to an injection energy of $E_{\text{inj}} \sim 50$ keV.
This is reasonably similar to the 60 keV injection energy required to fit the presented observations using the heated $\kappa$-distribution as a source.

Note the presented simulations are not intended to reproduce observations exactly, nor is it posited that the spectrum observed during the presented ESP event is the unambiguous result of particles accelerated from the SW.
The aforementioned results do however demonstrate that the prior heating of the SW plasma and energy distribution complements the shock acceleration process well and might even be necessary when considering shocks with large injection energies.  

\subsection{The evolution of shock-accelerated distributions during the shock passage} \label{subsec:spec_evolution}

\begin{figure}[tp]
\figurenum{17}
\centering
\includegraphics[scale=0.4]{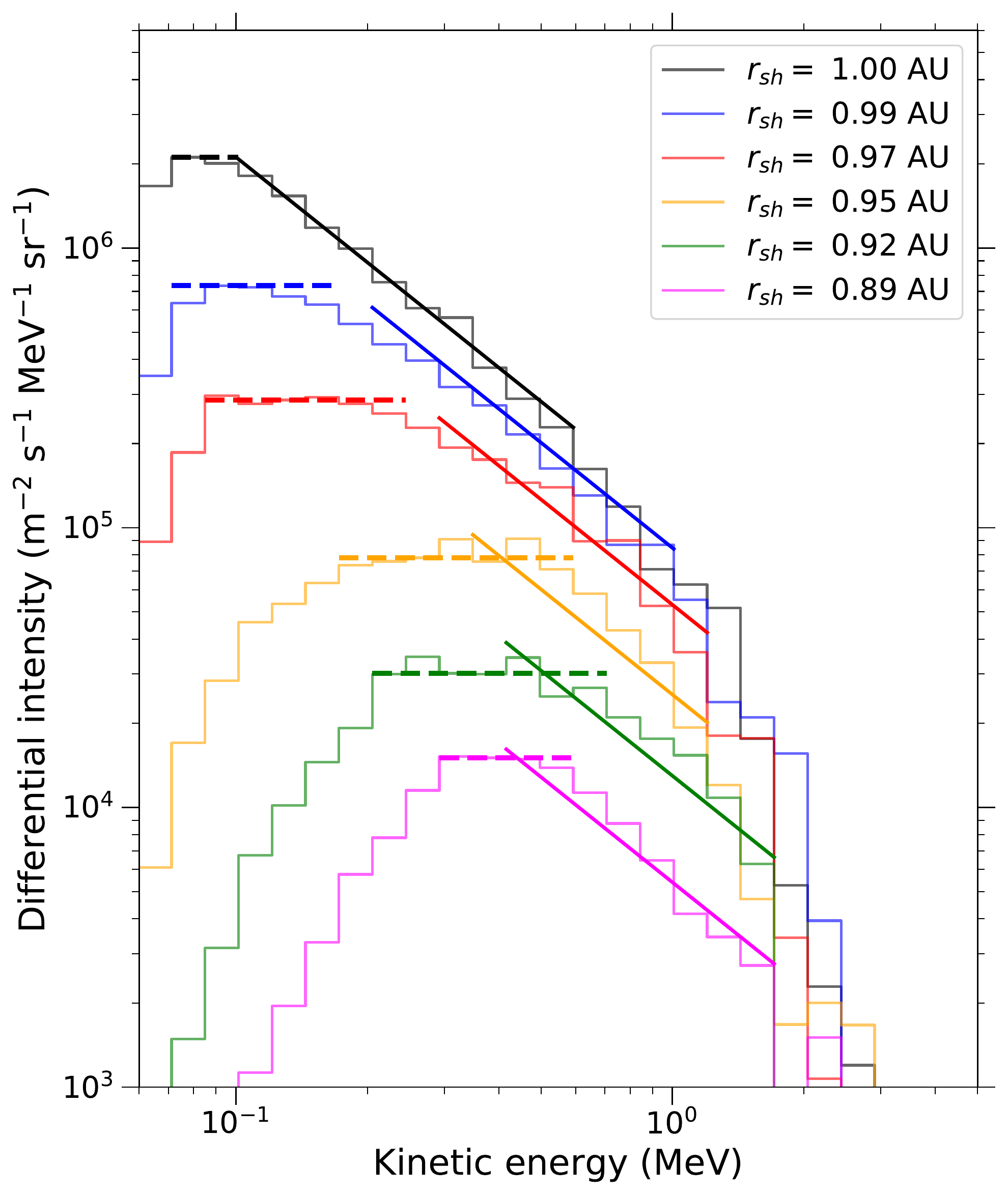}
\caption{Modelled shock-accelerated spectra at the Earth at different times during the approach of the shock. Step-like lines represent SDE solutions, dashed lines indicate flat segments, and the thick solid lines indicate power law segments that are distributed as $E^{-1.25}$ as expected from a shock with $s=3$.  \label{fig:spectra_shockapproach}}
\end{figure}

\begin{figure}[tp]
\figurenum{18}
\centering
\includegraphics[scale=0.4]{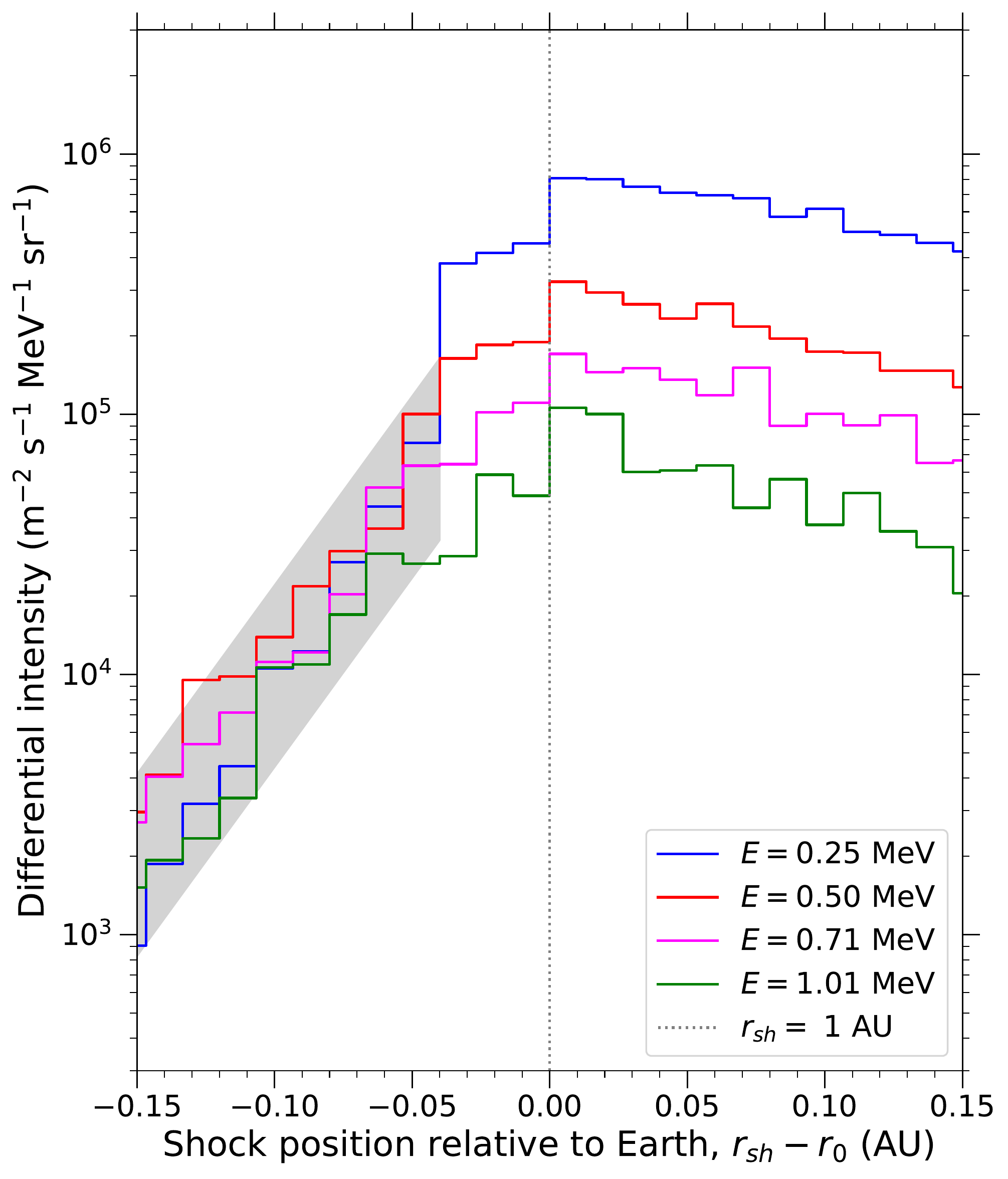}
\caption{Profiles of energetic particle intensities at different energies as viewed by an observer at Earth during the passage of the shock. The vertical dotted line represents the moment of the shock's arrival at Earth. The shaded region accentuates where energy spectra are approximately flat.  \label{fig:profiles_shockapproach}}
\end{figure}

Particles of sufficient energy can be injected into the DSA process at any time during a CME shock's passage between the Sun and the Earth.
However, ESP events entail the local enhancement of particle intensities as viewed by spacecraft at Earth.
The largest enhancement is naturally expected when the shock reaches Earth.
Solving the time-backward SDEs for the reference configuration and different starting positions of the shock, the energy spectra and intensity profiles of shock-accelerated particles are simulated for the approach and aftermath of the shock's passage at Earth.
These are shown in Figures \ref{fig:spectra_shockapproach} and \ref{fig:profiles_shockapproach}, respectively.  

The time-evolution of energy spectra during the approach of the shock is considered first.
From Figure \ref{fig:spectra_shockapproach}, the most recognizable aspect is of course the power-law form of the spectrum at the time of the shock's arrival at Earth, that is, where $r_{sh}=$ 1 AU, followed by a cut-off at higher energies.
However, this spectrum appears notably different when the shock is just 0.01 AU away, which for $V_{sh}=2400$ km s$^{-1}$ is a few minutes before its arrival at Earth.
The differences, which become more stark for larger distances between the shock and the Earth, include lower overall intensities and a downturn in the spectrum at low energies.
The progressively lower intensities follow merely as a result of the source of energetic particles (i.e. the shock) being further away.  
The low-energy downturns are more severe when the shock is far from the Earth, but manifest as flattened spectra for smaller distances from the Earth.
This likely follows because shock-accelerated particles are adiabatically cooled in the expanding SW while they propagate ahead of the shock toward Earth.
These steep spectra with positive power-law indices at lower energies are known spectral characteristics of adiabatic energy losses \citep[e.g.][]{MoraalPotgieter1982,Straussetal2011}.

A particularly interesting feature visible in Figure \ref{fig:spectra_shockapproach} is the flattening of spectra during the approach of the shock.
Similar flattening has recently been reported in observations of proton spectra between 50 keV and 1 MeV prior to the passage of an IP shock at Earth \citep{Larioetal2018}.
It is conceivable that this is the result of the competing effects of shock acceleration and adiabatic cooling of particles.
Whereas shock acceleration tends to distribute particles according to $E^{\gamma_s}$, where $\gamma_s=-1.25$ for $s=3$ according to Eq. \ref{eqn:specInd_DSA}, cooling tends to force protons into a characteristic $E^{+1}$ spectrum at Earth.
When the shock is nearer to Earth, the shock-accelerated component dominates, whereas particles transported from the shock while it is still further upwind have been cooled to a greater extent.
Where these effects balance, the spectrum flattens.
As a result of progressively stronger cooling, Figure \ref{fig:spectra_shockapproach} shows the flattened segments narrow and move to higher energies for larger distances between the shock and the Earth.
Figure \ref{fig:profiles_shockapproach} shows intensity profiles similar to how the observations of \cite{Larioetal2018} are presented. 
The flat energy spectra can be discerned from the coinciding intensity profiles of 0.25 to 1 MeV particles.
This energy range can be broadened in the simulations by specifying a lower injection energy or assuming a stronger shock, e.g. $s=4$, for which the shock-accelerated spectrum will be less steep.
These flat segments are visible for shock positions up to at least 0.15 AU away from the Earth, that is, for nearly three hours prior to its arrival.
Note the shock speed of 2400 km s$^{-1}$ implemented in these simulations is very fast. If a more typical shock speed of e.g. 1000 km s$^{-1}$ \citep[e.g.][]{Makelaetal2011} is assumed, this effect would be visible for a longer time prior to the shock's arrival at Earth.

The intensity profiles shown in Figure \ref{fig:profiles_shockapproach} largely resemble that of typical ESP events: a large onset of particle intensities over a relatively short time is visible before the arrival of the shock, followed by a more gradual decline of intensities after it has passed.
Note that the peaks of the simulated profiles, corresponding to the arrival of the IP shock at Earth, are not as sharp as observed intensities tend to show.
These peaks are often associated with large anisotropies \citep[e.g.][]{Larioetal2005a}, which can be more appropriately modelled using focused transport models \citep[][]{Zuoetal2011,leRouxWebb2012}.
The TPE in this study is solved for an omni-directional distribution function (as discussed in Appendix \ref{sec:countingparticles}) and describes only the near-isotropic particle component.

\subsection{Seed-particle energies and initial positions} \label{subsec:accsites}
 
\begin{figure}[tp]
\figurenum{19}
\centering
\includegraphics[scale=0.4]{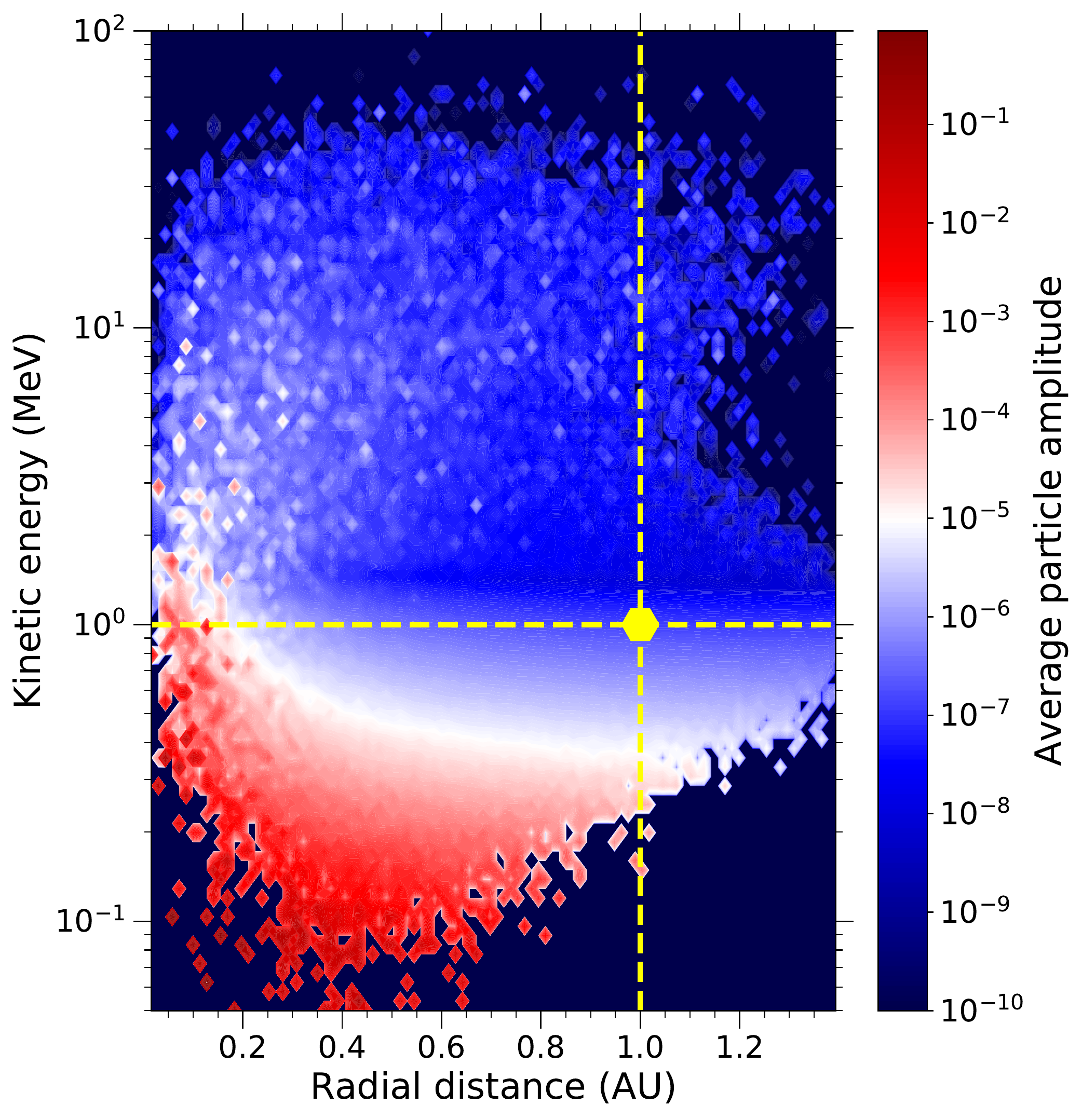}
\caption{Initial radial and kinetic energy distribution of SW particles contributing to 1 MeV intensities at Earth during an ESP event.
The color scale indicates the average particle amplitudes at each $(r,E)$-point, where the maximum amplitude has been normalised to unity.
The observational point, from where pseudo-particle trajectories are traced in a time-backwards fashion, is shown as a marker at the intersection of vertical and horizontal dashed lines indicating Earth's position and 1 MeV, respectively.
\textit{The online version shows the evolution of this plot as the shock progresses toward Earth, where the position of the shock front is indicated using a solid vertical line.} \label{fig:contours1MeV}}   
\end{figure}

It only remains to be investigated where and from which energies SW particles contributing to any given observational point are accelerated.
To do this, the binning technique discussed in Section \ref{subsec: SDEs} is implemented.
Firstly, pseudo particles are traced time-backwards from the observational point $(r_{\text{obs}},E_{\text{obs}}) = (\text{1 AU, 1 MeV})$, whereafter the average particle amplitudes are calculated for each $(r,E)$-bin at $t^{\prime}=0$.
As before, the simulation is run for the reference parameters listed in Table \ref{tab:shspeedV}.
The result is shown as a colour-scaled plot in Figure \ref{fig:contours1MeV}.

This plot essentially maps the relative contributions of particles from different initial positions and energies to the intensities of 1 MeV particles at Earth's position at the time of the shock's passage.
As such, they reveal a great deal of insight into the probable original energies and locations of seed particles.
Figure \ref{fig:contours1MeV} shows the undisturbed distribution (that is, before the departure of the shock from near to the Sun) of SW particles that would eventually contribute to 1 MeV intensities at Earth. 
It is illustrated that the largest contributions are from particles initially located upwind from Earth and accelerated from energies ranging from tens to a few hundred keV.
The lower limit of this range corresponds to the 60 keV injection energy implemented in this study.
The contributions of particles either cooled to 1 MeV from higher energies or convected to Earth without incurring energy changes are minuscule by comparison.
During the shock transit, the distribution of contributing particles will move closer to the observational point, until, at the time of the shock passage at Earth, only that point will be populated on the plot.  

The much larger relative contribution of particles accelerated to $E_{\text{obs}}$ from lower energies illustrates that DSA at the travelling IP shock is the chief contributor to intensities during the simulated ESP event. 

\section{Discussion and conclusions} \label{sec:conclusions}    

In this study, the acceleration of SW particles at halo CME-driven IP shocks is investigated.
The thermal and suprathermal components of the SW velocity (or equivalent energy) distributions are collectively described using $\kappa$-functions.
Furthermore, these SW distributions are transformed in response to simulated shock transitions in plasma properties upon which they depend, such as the flow speed, number density, and temperature.
These transformed (or heated) distributions are consequently specified as source spectra, from which particles with sufficient energy can be injected into the DSA process at the shock.
To model this acceleration process, SDEs equivalent to the Parker TPE are solved in a time-backwards fashion.
Using the combined approach of the pre-injection heating of the SW distribution and DSA, simulations reveal a number of noteworthy results with regards to the particle acceleration at IP shocks and particularly of SW particles. 

The simulations are shown to produce the classical spectral features of DSA.
However, in cases where diffusion length scales are small relative to the shock width, shock-accelerated spectra do not display the spectral indices associated with the full compression ratio.
Such situations may arise in the case of high levels of magnetic turbulence, which can decrease diffusion length scales, or in the case of broader compressions in the SW \citep[e.g.][]{Giacaloneetal2002}.
At any rate, the softer power-law distributions that are attained as a result notably reduces the overall intensities of shock-accelerated particles.
Furthermore, it is shown that the highest attainable energies of shock-accelerated spectra are limited by the transit time of the shock.
Reduced diffusion length scales may serve to better confine particles near the shock, thereby allowing spectra to be accelerated to higher energies within the available time.
Intensities of higher-energy particles ($E>$ 1 MeV) are shown to be particularly sensitive to this time limit.

The dependence of the acceleration process on shock properties such as its speed and compression ratio is also investigated.
Fast shocks contribute appreciably to the heating of the SW distribution.
While fast-moving shocks have shorter associated transit times, allowing less time for particle acceleration, larger shock speeds have a net positive effect on particle acceleration.
Through their contribution to larger flow speeds they reduce diffusion length scales, improving particle confinement at the shock.
They also reduce the time required for acceleration: 
since larger shock speeds imply larger differences of flow speeds across the shock, a particle scattered across it experiences a larger mean energy gain per crossing; the scattering centres can also be thought of converging on the shock at a higher rate for fast shocks.

The compression ratio, as measure of shock strength, also affects the magnitude of flow speed transitions across the shock, providing the actual factor by which both the shock-frame flow speed and number density change across the shock.
In particular, larger compression ratios increase intensities of the SW distribution across the shock, since the $\kappa$-function is normalised to the number density.
While it has a more modest effect on acceleration times than the shock speed, its effect on overall particle intensities is significant, since the spectral indices of the shock-accelerated spectra depend directly on the compression ratio.
When strong and fast shocks are compared as particle accelerators, it is found that strong slower-moving shocks produce larger numbers of energetic particles than weak fast-moving shocks.
The compression ratio, being directly associated with the steepness of accelerated energy distributions, is therefore identified to be the greater limiting factor between these two shock properties.

With regards to simulating ESP events, the prior heating of the SW distribution during the shock passage is found to complement the DSA process well:
it provides greater intensities of potential seed particles for DSA, especially where large injection energies are considered, and allows shock-accelerated spectra to achieve large enough intensities at sufficiently high energies to reproduce typical ESP events.
This result is consistent with observations reporting larger peak particle fluxes during ESP events for warmer CME sheath temperatures \citep{Dayehetal2018}.
Furthermore, simulations of shock-accelerated intensities at Earth reveal significant flattening of spectra forming ahead of the shock during its approach.
This is found to result from shock-accelerated particles experiencing adiabatic cooling in the expanding SW while propagating toward Earth ahead of the shock.
This provides a potential explanation for similar features recently reported in observations \citep{Larioetal2018}.

Finally, taking advantage of the time-backward tracing of phase-space density elements, with their flux contributions weighted according to particle amplitude, it is revealed that most SW particles contributing to intensities during a simulated ESP event are transported to Earth from upwind and are accelerated from energies ranging from tens to a few hundred keV.
Due to the outsized fraction of particles accelerated to 1 MeV from lower energies, it can be concluded that DSA is indeed a chief contributor to intensities at Earth during ESP events. 

It is ultimately concluded that with the combination of prior heating and shock acceleration, energetic particles can be accelerated from the suprathermal SW at IP shocks, and that these particles may make appreciable contributions to intensities during observed ESP events.
Also, the reproduction of the observed flat energy spectra ahead of shock passages \citep{Larioetal2018}, resulting from the competing processes of DSA and adiabatic cooling, illustrates the advantage of studying particle acceleration in association with more general transport processes.  
It is furthermore demonstrated that the SDE approach, while shown in Appendix \ref{sec:benchmark} to yield comparable results to finite-difference methods, can be used to reveal unique physical insights with regards to particle transport and acceleration at IP shocks.
Along with its computational advantages, the SDE model developed in this study can be expanded to include more spatial dimensions, providing a means to explore more complicated shock geometries.
Further research using this SDE approach may also include more refined characterisations of transport coefficients, the solution of focused transport equations to study large-anisotropy events, and the reproduction of various shock-related particle events, such as the acceleration of particles associated with co-rotating interactions regions.  

\acknowledgments PLP and RDS acknowledges the financial support of the South African National Research Foundation (NRF). Opinions expressed and conclusions arrived at are those of the authors and are not necessarily to be attributed to the NRF.  

\appendix

\section{Counting particles} \label{sec:countingparticles}

\subsection{The distribution function and number density} \label{subsec: def_f}

\begin{figure}[tp]
\figurenum{20}
\centering
\includegraphics[scale=0.4]{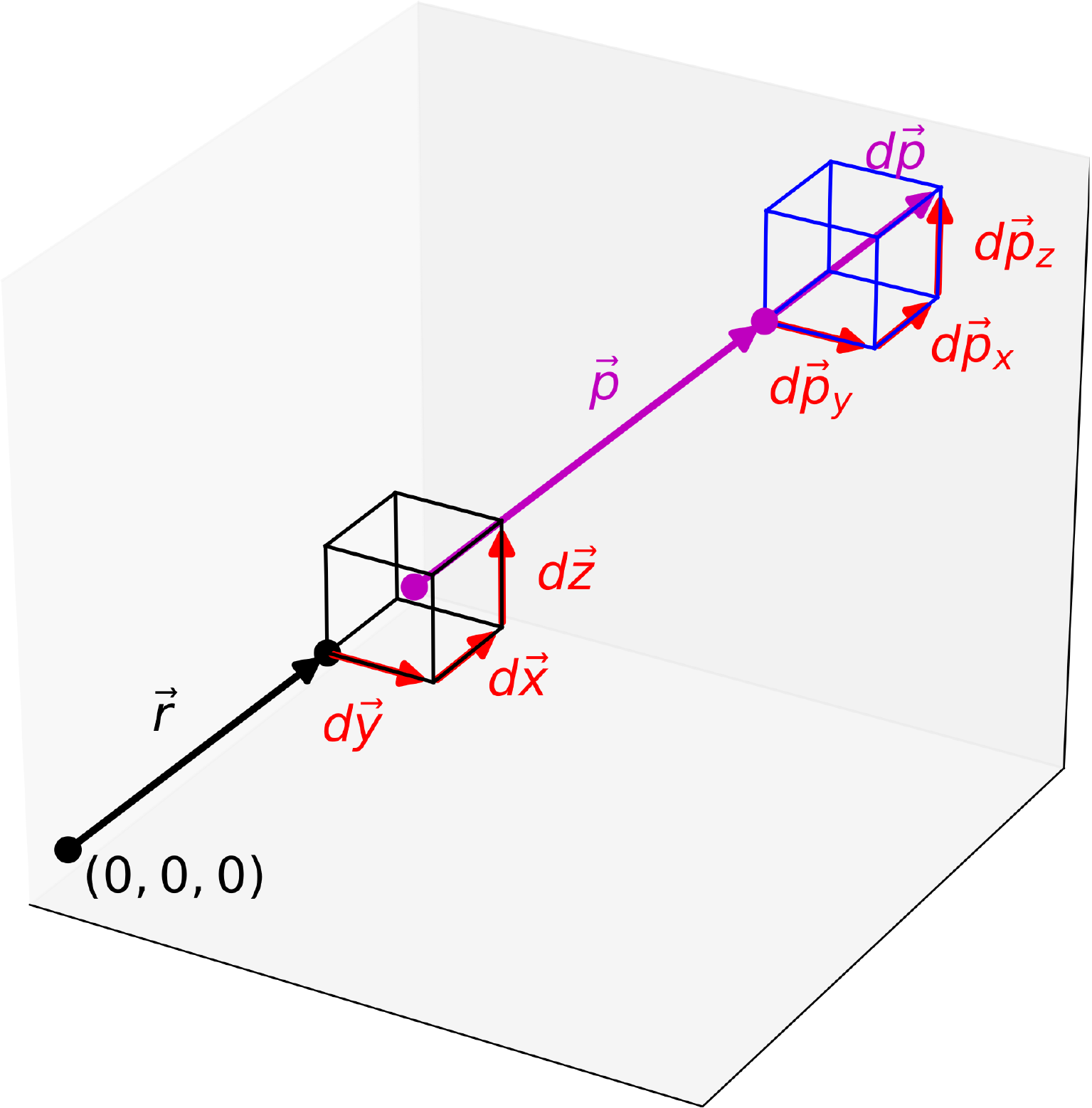}\hspace{3em} \includegraphics[scale=0.4]{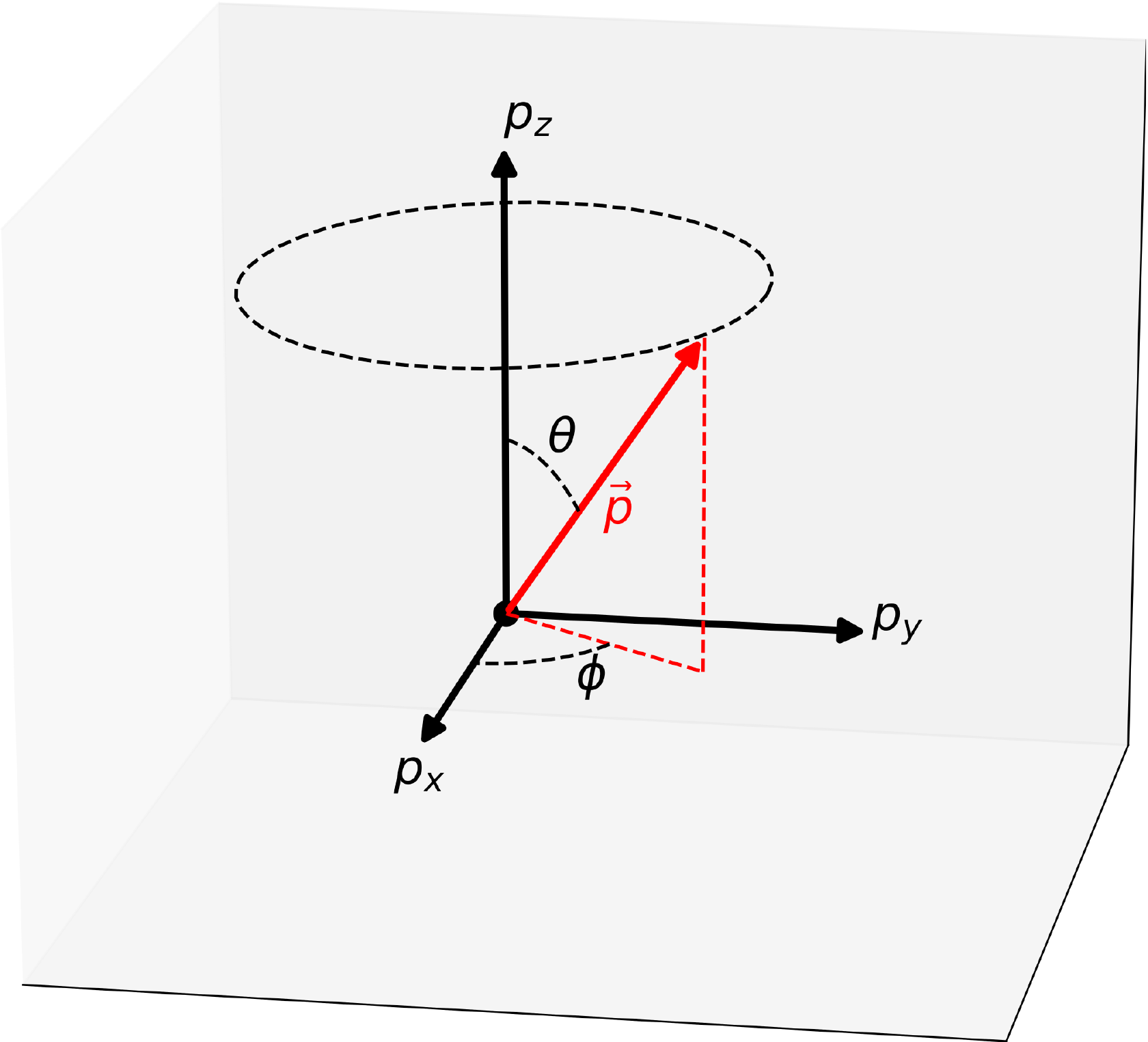}
\caption{Left: Representations of volumes in real and phase space, $d^3x$ and $d^3p$, for which the distribution function is defined according to Eq. \ref{eqn:distfunc_def}. $d^3x$ contains all the particles at position $\vec{r}$ with a momentum $\vec{p}$ in $d^3p$.  Right: A spherical momentum coordinate system, where $p_z$ is along the magnetic field line and the particle momentum $\vec{p}$ is defined in terms of the scalar momentum $p$, gyro-phase angle $\phi$ and pitch angle $\theta$.  
  \label{fig:def_dist_func}}
\end{figure}

Consider a number of particles $d \mathcal{N}$ in a small volume $d^3x$ at a position $\vec{x} = (x,y,z)$ and with momentum in $d^3p$ around $\vec{p} = (p_x, p_y, p_z)$ as illustrated in Figure \ref{fig:def_dist_func} (left). The distribution function $f\left(\vec{x},\vec{p},t\right)$ is introduced such that
	\begin{equation} \label{eqn:distfunc_def}
	d\mathcal{N} = f\left(\vec{x},\vec{p},t\right) d^3x\ d^3p \text{.} 
	\end{equation}
	This can be related to the ordinary number density $n$ by integrating over all momentum space,
	\begin{equation} \label{eqn:volint_f_numdense}
	 n = \int f\left(\vec{x},\vec{p},t\right) d^3p \text{.} 
	\end{equation}
The above volume integral can be re-written by considering a momentum coordinate system as shown in Figure \ref{fig:def_dist_func} (right). 
Assuming the gyro-centre at the origin and the magnetic field line about which the particle gyrates along $p_z$, that is $\vec{B}=B\hat{z}$, the azimuthal angle $\phi$ represents the particle's gyrophase and the polar angle $\theta$ its pitch angle.

This allows the momentum-space volume element to be expressed in spherical coordinates as $d^3p = p^2\ d\Omega\ dp$, where $d\Omega = \sin\theta\ d\theta\ d\phi$ is a solid-angle element about $\vec{p}$ and $p = |\ \vec{p}\ |$ is the particle's scalar momentum. Noting that $\vec{p}$ can hence be expressed in terms of the coordinates $(p,\ \phi,\ \theta)$, Eq. \ref{eqn:volint_f_numdense} can be rewritten as 
\begin{align} \label{eqn:volint_f_numdense_spherical}
	n &= \int\limits_0^{\infty} \int\limits_{0}^{2\pi} \int\limits_{0}^{\pi} p^2\ f\left(\vec{x},p,\theta,\phi,t\right) \sin\theta\ d\theta\ d\phi\ dp \nonumber \\
	&= \int\limits_0^{\infty}p^2\ \int\limits_{\Omega}  f\left(\vec{x},p,\theta,\phi,t\right) d\Omega\ dp \text{,}
\end{align}
where $\int_\Omega$ represents the integral over all solid angles.
However, since particle detectors typically cannot resolve statistically significant numbers of particles incident from only one particular direction, the average number of particles per unit solid angle in momentum space is considered. 
 Accordingly, the omni-directional distribution function (that is, averaged over all directions in $\vec{p}$-space for a fixed $p$) is defined as 
 \begin{equation} \label{eqn:omni_f}
	f_0\left(\vec{x},p,t\right) = \dfrac{\int\limits_{\Omega} f\left(\vec{x},\vec{p},t\right) d\Omega}{\int_{\Omega} d\Omega}	= \frac{1}{4\pi} \int\limits_{\Omega} f\left(\vec{x},\vec{p},t\right) d\Omega\  \text{,}
\end{equation}
since $\int_{\Omega} d\Omega = \int\limits_0^{2\pi} d\phi \int\limits_0^\pi \sin\theta d\theta = 4\pi$.
Note that $f_0\left(\vec{x},p,t\right)$ is the distribution function solved for in the \cite{Parker1965} transport equation. Note furthermore from Eq. \ref{eqn:omni_f} that $f\left(\vec{x},\vec{p},t\right)$ can be taken out of the integral over solid angle if it is assumed to be gyro- and isotropic, that is, independent of both $\phi$ and $\theta$. In this case it follows that $f\left(\vec{x},\vec{p},t\right)=f_0\left(\vec{x},p,t\right)$.
From Eqs. \ref{eqn:volint_f_numdense_spherical} and \ref{eqn:omni_f} the number density can be expressed as 
\begin{align} \label{eqn:omni_numdense}
 n &=	\int\limits_0^{\infty} 4\pi p^2\ f_0\left(\vec{x},p,t\right) dp \\ &= \int\limits_0^{\infty} U_{p}\left(\vec{x},p,t\right) dp \text{,} \label{eqn:omni_difdense}
\end{align}
where $U_{p}\left(\vec{x},p,t\right)=4\pi p^2\ f_0\left(\vec{x},p,t\right)$ is known as the differential number density. Here, $U_p$ represents the number density between two spherical shells in $\vec{p}$-space with radii of $p$ and $p+dp$, respectively.
While spacecraft cannot resolve volume densities, $U_p$ does prove useful to relate quantities they measure to those introduced above.

 \subsection{Toward the differential intensity} \label{subsec:toward_j}
  
Spacecraft essentially measure the flux of particles observed for a particular viewing direction within a particular momentum (or energy) band. 
This is known as the differential intensity $j_p$, and is the number of particles detected per momentum interval per unit time per unit area of observed space per unit solid angle.
Dimensionally, this can be attained by the product of the differential number density and the speed at which particles move toward the detector through a surface perpendicular to their motion, averaged over all solid angles. That is, 
\begin{align} \label{eqn:j_difnumdense_fo}
	j_p \left(\vec{x},p,t \right) &= \frac{v\ U_p\left(\vec{x},p,t \right)}{4\pi} \nonumber \\
	 &= v\ p^2\ f_0\left(\vec{x},p,t\right) \text{,}	
\end{align}
which, in turn, relates the differential intensity to the distribution function.
It is furthermore useful to express the differential intensity in terms of kinetic energy to compare with spacecraft observations.
The conversion of $j_p$, expressed per momentum interval $dp$, to $j_E$, expressed per interval of kinetic energy $dE$, is carried out by noting that $j_p\ dp = j_E\ dE$ due to the conservation of particles. 
It hence follows that
\begin{equation} \label{eqn:j_E_j_p}
j_E  = j_p  \frac{dp}{dE} \text{.} 
\end{equation}
Note that since $p^2 = \left( E_T^2 - E_0^2 \right)/c^2$, where $E_T = E + E_0$ with $E_T$, $E$ and $E_0$ denoting the total, kinetic and rest-mass energies, respectively, it follows that
\begin{equation} \label{eqn:dp_dE}
	\frac{dp}{dE}	= \frac{E+E_0}{pc^2} = \frac{E_T}{pc^2} = \frac{\gamma m_0 c^2}{\gamma m_0 v c^2}=\frac{1}{v} \text{,}
\end{equation}
where $\gamma$ is the Lorentz factor and $m_0$ is the rest mass. Then, using Eq. \ref{eqn:j_E_j_p}, $j_E$ can be related to the distribution function as follows
\begin{equation} \label{eqn:j_E_f0}
	j_E = \frac{j_p \left(\vec{x},p,t \right)}{v} = p^2\ f_0\left(\vec{x},p,t\right) \text{.}
\end{equation}

\subsection{On kappa velocity distribution functions} \label{subsec:on_kappa_funcs}

The normalization constant $A_{\kappa}$ of the standard $\kappa$-function introduced in Eq. \ref{eqn:kappafunc} is obtained by setting its integral over all phase space equal to the number density of the solar wind.  
Assuming an omni-directional (or three-dimensional) $\kappa$-function, averaged as per Eq. \ref{eqn:omni_f}, the expression given in Eq. \ref{eqn:omni_numdense} can be used to calculate the normalization constant. 
Note, however, because $f_{\kappa}$ is often expressed as a function of velocity instead of momentum, the analogous expression in velocity coordinates, namely
\begin{equation} \label{eqn:omni_numdense_v}
 n =	\int\limits_0^{\infty} 4\pi v^2\ f_{\kappa}\left(v\right) dv \text{,}
\end{equation}
is used to calculate the number density.
Bear in mind that for use in the transport equation, e.g. when specifying $f_s$ in the source function, the conversion $f_s = f_{\kappa}/m_p^3$ applies, because $\int 4\pi p^2\ f_s\ dp = \int 4\pi (m_p v)^2\ f_s\ d(m_p v) \equiv \int 4\pi v^2\ f_{\kappa}\ dv $.
Here, only the proton rest mass $m_p$ is used, since this study is concerned with solar wind protons with kinetic energies typically much smaller than their rest-mass energy.
A non-relativistic description is therefore applicable.
Note furthermore that Eq. \ref{eqn:omni_numdense_v} is also known as the zeroth-order velocity moment of $f_{\kappa}$, and that it can be cast into a more general form for the $n^{\text{th}}$-order moment, namely
\begin{equation} \label{eqn:nth_order_moment}
	\avg{v^n} = \int\limits_0^{\infty} 4\pi v^{n+2}\ f_{\kappa}\left(v\right) dv  \text{,}
\end{equation}
which, upon substituting Eq. \ref{eqn:kappafunc}, becomes
\begin{equation} \label{eqn:nth_order_kappa}
	\avg{v^n} = 4\pi A_{\kappa} \int\limits_0^{\infty} v^{n+2}\ \left( 1+\frac{v^2}{\kappa v_{\kappa}^2} \right)^{-(\kappa+1)} dv   \text{,}
\end{equation}
from which the normalization constant $A_{\kappa}$ can be calculated. 
The integral in Eq. \ref{eqn:nth_order_kappa} is evaluated by applying the transformation $v^2 / (\kappa v_{\kappa}^2) :\rightarrow \eta$, which allows the expression to be rewritten as 
\begin{equation} \label{eqn:nth_order_kappa_transformed}
	\avg{v^n} = 2\pi A_{\kappa}\ (\kappa v_{\kappa}^2)^{(n+3)/2}  \int\limits_0^{\infty} \eta^{(n+1)/2}\ \left( 1+\eta \right)^{-(\kappa+1)} d\eta   \text{.}
\end{equation}
Note that the integral is now in the form of the $\beta$-function, given by
\begin{equation} \label{eqn:beta_function}
\beta(x,y) = \int\limits_0^{\infty} \eta^{x-1}\ (1+\eta)^{-(x+y)} d\eta = \frac{\Gamma(x)\Gamma(y)}{\Gamma(x+y)} \text{.}
\end{equation}
Recognizing that $x = (n+3)/2$, $y = \kappa - (n+1)/2$, and $x+y = \kappa +1$, the integral in Eq. \ref{eqn:nth_order_kappa_transformed} can be solved by invoking the identity given in Eq. \ref{eqn:beta_function}, so that
\begin{equation} \label{eqn:nth_order_kappa_solved}
	\avg{v^n} = 2\pi A_{\kappa}\ (\kappa v_{\kappa}^2)^{(n+3)/2}\ \frac{\Gamma((n+3)/2)\Gamma(\kappa - (n+1)/2)}{\Gamma(\kappa +1)} \text{,}
\end{equation}
from which the normalization constant can be determined as
\begin{equation} \label{eqn:A_kappa}
A_{\kappa} = \frac{n_{sw}}{2\pi(\kappa v_{\kappa}^2)^{3/2}}\ \frac{\Gamma(\kappa+1)}{\Gamma(3/2)\Gamma(\kappa-1/2)}
\end{equation}
by setting $\avg{v^{n=0}}=n_{sw}$, yielding the same expression given in Eq. \ref{eqn:kappanormconst}. The standard kappa function is hence given by
\begin{equation} \label{eqn:A_kappa_final}
f_{\kappa} = \frac{n_{sw}}{2\pi(\kappa v_{\kappa}^2)^{3/2}}\ \frac{\Gamma(\kappa+1)}{\Gamma(3/2)\Gamma(\kappa-1/2)}\ \left( 1+\frac{v^2}{\kappa v_{\kappa}^2} \right)^{-\kappa -1} \text{.}
\end{equation}

\section{Equivalence of SDE and finite-difference numerical methods: Application to an event during the 2003 Halloween epoch} \label{sec:benchmark}

In a similar study, where DSA is also assumed as an acceleration mechanism in simulating ESP events, \cite{Giacalone2015} utilizes a finite-difference numerical scheme to solve the \cite{Parker1965} TPE in a single spatial dimension for a strong fast-moving shock ($V_{sh}=$ 1900 km s$^{-1}$, $s=4$).
In that study, the energetic particle event, or \textit{Halloween} event, of 2003 October 29 is considered as an application for which observed energy spectra and temporal profiles are reproduced.
To demonstrate their equivalence as numerical methods, the SDE approach of this study is implemented to reproduce particle intensities for the same event using a similar parameter configuration to that of \cite{Giacalone2015}.
The source function in that study is bound to a numerical grid, which does not exist in the SDE approach. 
For the purposes of this application, the source function is instead specified at the shock as a very soft power law in momentum ($p^{-7}$), where simulated intensities are retroactively normalised to observed intensities.    
Furthermore, the diffusion coefficient is implemented as specified in that study, that is, of order $10^{19}$ cm$^2$ s$^{-1}$ near the Sun and with a momentum dependence of $\sim p^{1.5}$ \citep[refer to][for further details]{Giacalone2015}.

\begin{figure*}[tp]
\figurenum{21}
\centering
\includegraphics[scale=0.4]{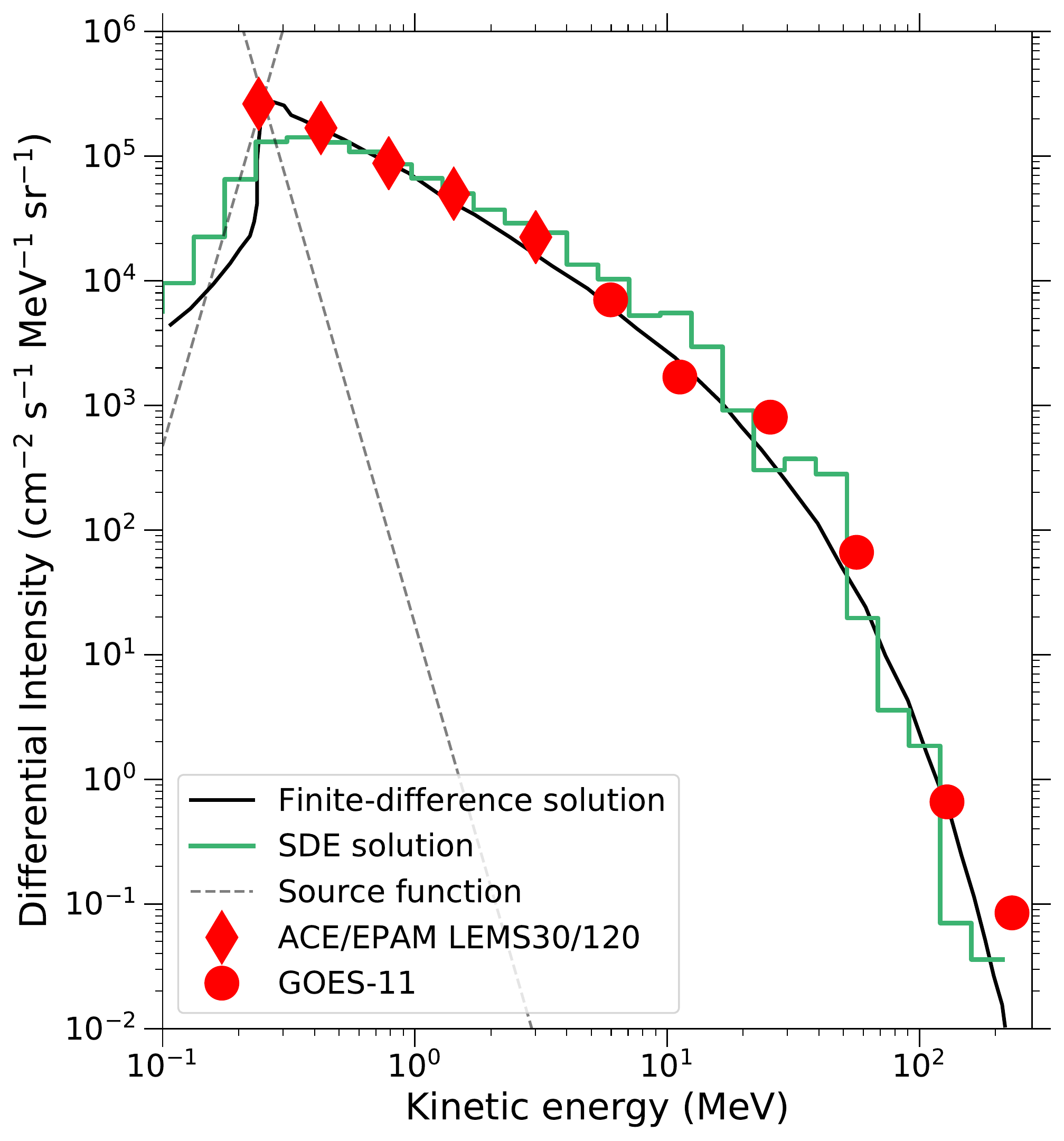}\ \includegraphics[scale=0.4]{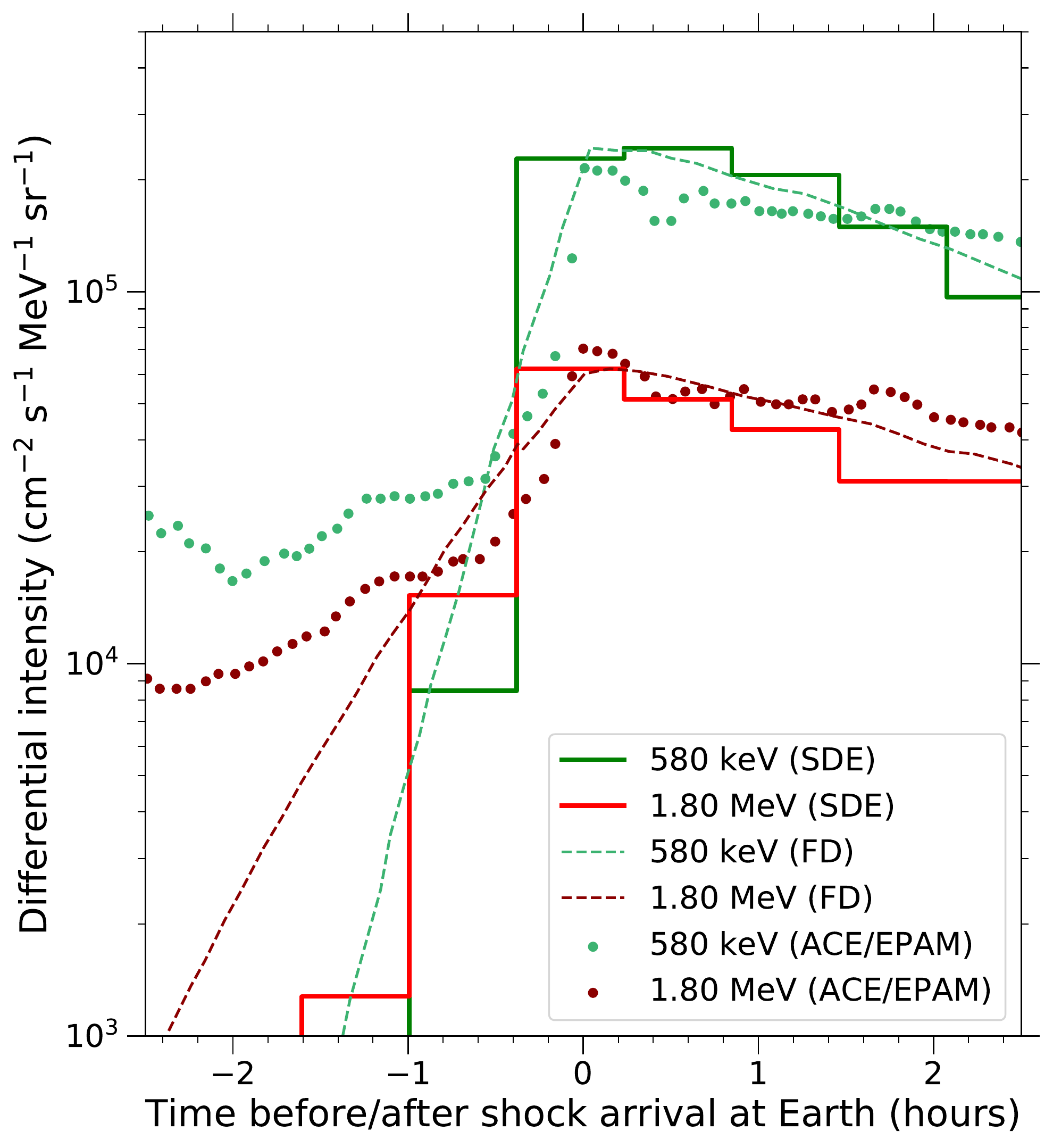}
\caption{ Left: Energy spectra at Earth's position at the time of a CME-driven shock passage on 2003 October 29. Using similar parameter configuations, solutions of the SDE approach used in this study and the finite-difference method of \cite{Giacalone2015} are shown in solid green step-like and black lines, respectively. The red diamond and circle markers represent observations from  \textit{ACE}/EPAM (LEMS30 and LEMS120) and \textit{GOES-11}, respectively. Right: Temporal profiles of 580 keV and 1.8 MeV ion intensities as observed at Earth during the same event considered on the left. The SDE and finite-difference solutions are shown as solid step-like and dashed lines, respectively, while dot-markers indicate observations from \textit{ACE}/EPAM (LEMS30 and LEMS120). See \cite{Giacalone2015} and the references therein for details.    \label{fig:halloween}}
\end{figure*}

The resultant energy spectra and time profiles of energetic particle intensities are shown in Figure \ref{fig:halloween}.
Overall, the finite-difference and SDE solutions are similar. 
Recall that the SDEs must be solved for a large number of pseudo particles for each energy, position or point in time at which intensities are sought.
The resolution of SDE solutions therefore merely depends on the number of times the simulation is repeated and how the observational points are distributed, as opposed to the fixed grid resolution of finite-difference schemes.
For example, in Figure \ref{fig:halloween}, a greater number of more tightly spaced observational points are chosen in energy than in time, as shown in the left and right-hand side panels, respectively.
Note that the SDE solutions tend to undershoot the finite-difference solutions where steeper gradients exist. 
This follows because there is no passage of information between pseudo particles and their associated observational points as would exist between adjacent points in finite-difference schemes. 
The SDEs typically yield less gradual intensity gradients as a result. 
This is useful to bear in mind when fitting functions to shock-accelerated spectra simulated using these two different approaches, since the high-energy cut-off is likely to be more abrupt for the SDEs than for finite-difference solutions.

Note, as an aside, that simulated intensities of either method reproduce observations better near the shock than away from it. 
This follows since the only source of particles in the simulations is situated on the shock itself, while the observed intensities include contributions from other sources.
In any event, the comparison of the simulations generated using the two different numerical methods demonstrate the equivalence of their results.

\bibliographystyle{aasjournal.bst}

\begin{thebibliography}{}
\expandafter\ifx\csname natexlab\endcsname\relax\def\natexlab#1{#1}\fi

\bibitem[{{Arthur} \& {le Roux}(2013)}]{ArthurLeRoux2013}
{Arthur}, A.~D., \& {le Roux}, J.~A. 2013, \apjl, 772, L26

\bibitem[{{Axford} {et~al.}(1977)}]{Axfordetal1977}
{Axford}, W.~I., {Leer}, E., \& {Skadron}, G. 1977, ICRC, 11, 132

\bibitem[{{Bell} (1978)}]{Bell1978a}
{Bell}, A.~R. 1978, \mnras, 182, 147

\bibitem[{{Bell} (1978)}]{Bell1978b}
{Bell}, A.~R. 1978, \mnras, 182, 443

\bibitem[{{Blandford} \& {Ostriker}(1978)}]{BlandfordOstriker1978}
{Blandford}, R.~D. and {Ostriker}, J.~P. 1978, \apjl, 221, L29

\bibitem[{{Bryant} {et~al.}(1962)}]{Bryantetal1962}
{Bryant}, D.~A., {Cline}, T.~L., {Desai}, U.~D., \& {McDonald}, F.~B. 1962, \jgr, 67, 4983

\bibitem[{{Caprioli} {et~al.}(2015)}]{Capriolietal2015}
{Caprioli}, D., {Pop}, A.~R., \& {Spitkovsky}, A. 2015, \apjl, 798, L28

\bibitem[{{Channok} {et~al.}(2005)}]{Channoketal2005}
{Channok}, C., {Ruffolo}, D., {Desai}, M.~I., \& {Mason}, G.~M. 2005, \apjl, 633, L53

\bibitem[{{Chateau} \& {Meyer-Vernet}(1991)}]{ChateauMeyerVernet1991}
{Chateau}, Y.~F., \& {Meyer-Vernet}, N. 1991, \jgr, 96, 5825

\bibitem[{{Chotoo} {et~al.}(2000)}]{Chotooetal2000}
{Chotoo}, K., {Schwadron}, N.~A., {Mason}, G.~M., {et~al.} 2000, \jgr, 105, 23107

\bibitem[{{Collier} {et~al.}(1996)}]{Collieretal1996}
{Collier}, M.~R., {Hamilton}, D.~C., {Gloeckler}, G., {Bochsler}, P., \& {Sheldon}, R.~B. 1996, \grl, 23, 1191

\bibitem[{{Desai} \& {Giacalone}(2016)}]{DesaiGiacalone2016}
{Desai}, M.~I., \& {Giacalone}, J. 2016, LRSP, 13, 3

\bibitem[{{Desai} {et~al.}(2006)}]{Desaietal2006}
{Desai}, M.~I., {Mason}, G.~M., {Mazur}, J.~E., \& {Dwyer}, J.~R. 2006, \ssr, 124, 261

\bibitem[{{Dayeh} {et~al.}(2018)}]{Dayehetal2018}
{Dayeh}, M.~A, {Desai}, M.~I., {Ebert}, R.~W., {et~al.} 2018, J. Phys. Conf. Ser., 1100, 012008

\bibitem[{{Drury} (1983)}]{Drury1983}
{Drury}, L.~O. 1983, Rept. Progr. Phys., 46, 973

\bibitem[{{Ellison} \& {Ramaty}(1985)}]{EllisonRamaty1985}
{Ellison}, D.~C. and {Ramaty}, R. 1985, \apj, 298, 400

\bibitem[{{Ellison} {et~al.}(1990)}]{Ellisonetal1990}
{Ellison}, D.~C., {Jones}, F.~C., \& {Reynolds}, S.~P. 1990, \apj, 360, 702

\bibitem[{{Ellison} {et~al.}(1995)}]{Ellisonetal1995}
{Ellison}, D.~C., {Baring}, M.~G., \& {Jones}, F.~C. 1995, \apj, 453, 873

\bibitem[{{Formisano} {et~al.}(1973)}]{Formisanoetal1973}
{Formisano}, V., {Moreno}, G., {Palmiotto}, F., \& {Hedgecock}, P.~C. 1973, \jgr, 78, 3714

\bibitem[{{Giacalone} (2005)}]{Giacalone2005}
{Giacalone}, J. 2005, \apjl, 628, L37

\bibitem[{{Giacalone} (2012)}]{Giacalone2012}
{Giacalone}, J. 2012, \apj, 761, 28

\bibitem[{{Giacalone} (2015)}]{Giacalone2015}
{Giacalone}, J. 2015, \apj, 799, 80

\bibitem[{{Giacalone} {et~al.}(1992)}]{Giacaloneetal1992}
{Giacalone}, J., {Burgess}, D., {Schwartz}, S.~J., \& {Ellison}, D.~C. 1992, \grl, 19, 433

\bibitem[{{Giacalone} \& {Jokipii}(1999)}]{GiacaloneJokipii1999}
{Giacalone}, J., \& {Jokipii}, J.~R. 1999, \apj, 520, 204

\bibitem[{{Giacalone} {et~al.}(2002)}]{Giacaloneetal2002}
{Giacalone}, J., {Jokipii}, J.~R., \& {K{\'o}ta}, J. 2002, \apj, 573, 845

\bibitem[{{Gopalswamy} {et~al.}(2005)}]{Gopalswamyetal2005}
{Gopalswamy}, N., {Yashiro}, S., {Liu}, Y., {et~al.} 2005, \jgr, 110, A09S15

\bibitem[{{Hellberg} {et~al.}(2009)}]{Hellbergetal2009}
{Hellberg}, M.~A., {Mace}, R.~L., {Baluku}, T.~K., {Kourakis}, I., \& {Saini}, N.~S. 2009, Phys. Plasmas, 16, 094701

\bibitem[{{Ho} {et~al.}(2009)}]{Hoetal2009}
{Ho}, G.~C. and {Lario}, D. and {Decker}, R.~B. 2009, AIP Conf. Proc, 1183, 19

\bibitem[{{Hu} {et~al.}(2017)}]{Huetal2017}
{Hu}, J., {Li}, G., {Ao}, X., {Zank}, G.~P., \& {Verkhoglyadova}, O. 2017, \jgr, 122, 10

\bibitem[{{Huttunen-Heikinmaa} \& {Valtonen}(2009)}]{HuttunenHeikinmaaValtonen2009}
{Huttunen-Heikinmaa}, K. \& {Valtonen}, E. 2009, AnGeo, 27, 767

\bibitem[{{Jokipii} (1966)}]{Jokipii1966}
{Jokipii}, J.~R. 1966, \apj, 146, 480

\bibitem[{{Jones} \& {Ellison}(1991)}]{JonesEllison1991}
{Jones}, F.~C., \& {Ellison}, D.~C. 1991, \ssr, 58, 259

\bibitem[{{Kallenrode} (1993)}]{Kallenrode1993}
{Kallenrode}, M.-B. 1993, \jgr, 98, 19

\bibitem[{{Kallenrode} {et~al.}(1992)}]{Kallenrodeetal1992}
{Kallenrode}, M.-B., {Wibberenz}, G., \& {Hucke}, S. 1992, \apj, 394, 351

\bibitem[{{Kang} {et~al.}(2014)}]{Kangetal2014}
{Kang}, H., {Petrosian}, V., {Ryu}, D., \& {Jones}, T.~W. 2014, \apj, 788, 142

\bibitem[{{Klein} \& {Dalla}(2017)}]{KleinDalla2017}
{Klein}, K.~L., \& {Dalla}, S. 2017, \ssr, 212, 1107

\bibitem[{{Kong} {et~al.}(2017)}]{Kongetal2017}
{Kong}, F.-J., {Qin}, G., {Zhang}, L.-H. 2017, \apj, 845, 43

\bibitem[{{Kr{\"u}lls} \& {Achterberg}(1994)}]{KruellsAchterberg1994}
{Kr{\"u}lls}, W.~M \& {Achterberg}, A. 1994, \aap, 286, 314

\bibitem[{{Krymskii} (1977)}]{Krymsky1977}
{Krymskii}, G.~F. 1977, DoSSR, 234, 1306

\bibitem[{{Lario} \& {Decker}(2002)}]{LarioDecker2002}
{Lario}, D. \& {Decker}, R.~B. 2002, \grl, 29, 1393

\bibitem[{{Lario} {et~al.}(2005a)}]{Larioetal2005a}
{Lario}, D., {Decker}, R.~B., {Livi}, S., {et~al.} 2005a, \jgr, 110, A09S11

\bibitem[{{Lario} {et~al.}(2005b)}]{Larioetal2005b}
{Lario}, D., {Hu}, Q., {Ho}, G.~C., {Decker}, R.~B., {Roelof}, E.~C., \& {Smith}, C.~W. 2005b, ESASP, 592, 81L

\bibitem[{{Lario} {et~al.}(2018)}]{Larioetal2018}
{Lario}, D., {Berger}, L., {Wilson}, L.~B. III, {et~al.} 2018, J. Phys. Conf. Ser., 1100, 012014

\bibitem[{{Lee} (1983)}]{Lee1983}
{Lee}, M.~A. 1983, \jgr, 88, 6109

\bibitem[{{Leubner} (2004)}]{Leubner2004}
{Leubner}, M.~P. 2004, Phys. Plasmas, 11, 1308

\bibitem[{{le Roux} \& {Arthur}(2017)}]{leRouxArthur2017}
{le Roux}, J.~A. \& {Arthur}, A.~D. 2017, J. Phys.: Conf. Ser., 900, 012013

\bibitem[{{le Roux} {et~al.}(1996)}]{leRouxetal1996}
{le Roux}, J.~A., {Potgieter}, M.~S., \& {Ptuskin}, V.~S. 1996, \jgr, 101, 4791

\bibitem[{{le Roux} \& {Webb}(2012)}]{leRouxWebb2012}
{le Roux}, J.~A. \& {Webb}, G.~M. 2012, \apj, 746, 104

\bibitem[{{Li} (2017)}]{Li2017}
{Li}, G. 2017, Science China Earth Sciences, doi:  10.1007/s11430-017-9083-y

\bibitem[{{Livadiotis} (2015)}]{Livadiotis2015}
{Livadiotis}, G. 2015, \apj, 809, 111

\bibitem[{{Livadiotis} \& {McComas}(2013)}]{LivadiotisMcComas2013}
{Livadiotis}, G., \& {McComas}, D.~J. 2013, \ssr, 175, 183

\bibitem[{{M{\"a}kel{\"a}} {et~al.}(2011)}]{Makelaetal2011}
{M{\"a}kel{\"a}}, P., {Gopalswamy}, N., {Akiyama}, S., {Xie}, H., \& {Yashiro}, S. 2011, \jgr, 116, A08101

\bibitem[{{Maksimovic} {et~al.}(1997)}]{Maksimovicetal1997}
{Maksimovic}, M., {Pierrard}, V., \& {Riley}, P. 1997, \grl, 24, 1151

\bibitem[{{Malkov} \& {V{\"o}lk}(1995)}]{MalkovVoelk1995}
{Malkov}, M.~A., \& {V{\"o}lk}, H.~J. 1995, \aap, 300, 605

\bibitem[{{Marcowith} \& {Kirk}(1999)}]{MarcowithKirk1999}
{Marcowith}, A. \& {Kirk}, J.~G. 1999, \aap, 347, 391

\bibitem[{{Maruyama} (1955)}]{Maruyama1955}
{Maruyama}, G. 1955, Rend. Circolo Mat. Palermo, 4, 48

\bibitem[{{Moloto} {et~al.}(2019)}]{Molotoetal2018}
{Moloto}, K.~D., {Engelbrecht}, N.~E., {Strauss}, R.~D., {Moeketsi}, D.~M., \& {van den Berg}, J.~P. 2019, Adv. Space Res., 63, 626

\bibitem[{{Moraal} \& {Potgieter}(1982)}]{MoraalPotgieter1982}
{Moraal}, H. \& {Potgieter}, M.~S. 1982, \apss, 84, 519

\bibitem[{{Mostafavi} {et~al.}(2017)}]{Mostafavietal2017}
{Mostafavi}, P., {Zank}, G.~P., \& {Webb}, G.~M. 2017, \apj, 841, 4

\bibitem[{{Neergaard Parker} \& {Zank}(2012)}]{NeergaardParkerZank2012}
{Neergaard Parker}, L., \& {Zank}, G.~P. 2012, \apj, 757, 97

\bibitem[{{Neergard Parker} {et~al.}(2014)}]{NeergaardParkeretal2014}
{Neergaard Parker}, L., {Zank}, G.~P., \& {Hu}, Q. 2014, \apj, 782, 52

\bibitem[{{Parker} (1958)}]{Parker1958}
{Parker}, E.~N 1958, \apj, 128, 664

\bibitem[{{Parker} (1965)}]{Parker1965}
{Parker}, E.~N 1965, \planss, 13, 9

\bibitem[{{Pierrard} \& {Lazar}(2010)}]{PierrardLazar2010}
{Pierrard}, V., \& {Lazar}, M. 2010, \solphys, 267, 153

\bibitem[{{Pei} {et~al.}(2010)}]{Peietal2010}
{Pei}, C., {Bieber}, J.~W., {Burger}, R.~A., \& {Clem}, J. 2010, \jgr, 115, A12107

\bibitem[{{Qureshi} {et~al.}(2003)}]{Qureshietal2003}
{Qureshi}, M.~N.~S., {Pallocchia}, G., {Bruno}, R., {et~al.} 2003, AIP Conf. Proc, 679, 489

\bibitem[{{Richardson} {et~al.}(2005)}]{Richardsonetal2005}
{Richardson}, J.~D., {Wang}, C., {Kasper}, J.~C, \& {Liu}, Y. 2005, \grl, 32, L03S03


\bibitem[{{Sapunova} {et~al.}(2017)}]{Sapunovaetal2017}
{Sapunova}, O.~V., {Borodkova}, N.~L., {Eselevich}, V.~G., {Zastenker}, G.~N., \& {Yermolaev}, Y.~I. 2017, Cosmic Research, 55, 396

\bibitem[{{Skoug} {et~al.}(2004)}]{Skougetal2004}
{Skoug}, R.~M., {Gosling}, J.~T., {Steinberg}, J.~T., {et~al.} 2004, \jgr, 109, A09102

\bibitem[{{Steenberg} \& {Moraal}(1999)}]{SteenbergMoraal1999}
{Steenberg}, C.~D, \& {Moraal}, H. 1999, \jgr, 104, 24879

\bibitem[{{Strauss} \& {Effenberger}(2017)}]{StraussEffenberger2017}
{Strauss}, R.~D., \& {Effenberger}, F. 2017, \ssr, 212, 151

\bibitem[{{Strauss} {et~al.}(2011)}]{Straussetal2011}
{Strauss}, R.~D., {Potgieter}, M.~S., {Kopp}, A., \& {B{\"u}sching}, I., 2011, \jgr, 116, A12105

\bibitem[{{Strauss} {et~al.}(2013)}]{Straussetal2013}
{Strauss}, R.~D., {Potgieter}, M.~S., {Ferreira}, S.~E.~S., {Fichtner}, H., \& {Scherer}, K. 2013, \apjl, 765, L18

\bibitem[{{Sundberg} {et~al.}(2016)}]{Sundbergetal2016}
{Sunberg}, T., {Haynes}, C.~T., {Burgess}, D., \& {Mazelle}, C.~X. 2013, \apj, 820, 21

\bibitem[{{Vasyliunas} (1968)}]{Vasyliunas1968}
{Vasyliunas}, V.~M. 1968, \jgr, 73, 2839

\bibitem[{{Wu} {et~al.}(2005)}]{Wuetal2005}
{Wu}, C.-C., {Wu}, S.~T., {Dryer}, M., {et~al.} 2005, \jgr, 110, A09S17

\bibitem[{{Zank} (2017)}]{Zank2017}
{Zank}, G.~P. 2017, in Kappa Distributions: Theory and Applications in Plasmas (Elsevier), 609

\bibitem[{{Zank} {et~al.}(2006)}]{Zanketal2006}
{Zank}, G.~P., {Li}, G., {Florinski}, V., {et~al.} 2006, \jgr, 111, A06108

\bibitem[{{Zhang} (1999)}]{Zhang1999}
{Zhang}, M. 1999, \apj, 513, 409

\bibitem[{{Zhang} (2000)}]{Zhang2000}
{Zhang}, M. 2000, \apj, 541, 428

\bibitem[{{Zuo} {et~al.}(2011)}]{Zuoetal2011}
{Zuo}, Z., {Zhang}, M., {Gamayunov}, K., {Rassoul}, H., \& {Luo}, X. 2011, \apj, 738, 168

\end{thebibliography}

\end{document}